# SALTUS Probe Class Space Mission: Observatory Architecture and Mission Design

**Leon K. Harding,[a,*] Jonathan W. Arenberg,[a] Benjamin Donovan,[a] Dave Oberg,[a] Ryan Goold,[a] Bob Chang,[a] Christopher Walker,[b] Dana Turse,[c] Jim Moore,[d] Jim C. Pearson Jr,[d] John N. Kidd Jr,[e] Zach Lung,[f] Dave Lung[f]**

[a]Northrop Grumman, 45101 Warp Drive, Dulles, Virginia 20166, United States
[b]Department of Astronomy, University of Arizona 933 N Cherry Ave., Tucson, Arizona 85721, United States
[c]Redwire Space Solutions, LLC, 2602 Clover Basin Drive, Longmont, Colorado 80503, United States
[d]NeXolve, 290 Dunlop Blvd, Building 200, Huntsville, Alabama 35824, United States
[e]Ascending Node Technologies, LLC, 2548 E 4th Street, Tucson, Arizona 85716, United States
[f]DA2 Ventures, 3970 Ridgeline Drive, Timnath, Colorado 80547, United States

**Abstract**. We describe the space observatory architecture and mission design of the *Single Aperture Large Telescope for Universe Studies (SALTUS)* mission, a NASA Astrophysics Probe Explorer concept. *SALTUS* will address key far-infrared science using a 14-m diameter <45 K primary reflector (M1) and will provide unprecedented levels of spectral sensitivity for planet, solar system, and galactic evolution studies, and cosmic origins. Drawing from Northrop Grumman's extensive NASA mission heritage, the observatory flight system is based on the LEOStar-3 spacecraft platform to carry the *SALTUS* Payload. The Payload is comprised of the inflation control system (ICS), Sunshield Module (SM), Cold Corrector Module (CCM), Warm Instrument Electronics Module, and Primary Reflector Module (PRM). The 14-m M1 is an off-axis inflatable membrane radiatively cooled by a two-layer sunshield (~1,000 m² per layer). The CCM corrects for residual aberration from M1 and delivers a focused beam to two instruments – the High Resolution Receiver (HiRX) and SAFARI-*Lite*. The CCM and PRM reside atop a truss-based composite deck which also provides a platform for the attitude control system. The *SALTUS* 5-year mission lifetime is driven by a two-consumable architecture: the propellant system and the ICS. The Core Interface Module (CIM), a multi-faceted composite truss structure, provides a load path with high stiffness, mechanical attachment, and thermal separation between the Payload and spacecraft. The SM attaches outside the CIM with its aft end integrating directly to the bus. The spacecraft maintains an attitude off M1's boresight with respect to the Sun line to facilitate the <45 K thermal environment. *SALTUS* will reside in a Sun-Earth halo L2 orbit with a maximum Earth slant range of 1.8 million km thereby reducing orbit transfer delta-v. The instantaneous field of regard provides two continuous 20º viewing zones around the ecliptic poles resulting in full sky coverage in six months.

**Keywords**: SALTUS, spacecraft, space observatory, mission architecture, deployable, sunshield.

*****Leon K. Harding**, E-mail: leon.harding@ngc.com

## 1   Introduction

Northrop Grumman has developed an observatory concept and mission design for the *Single Aperture Large Telescope for Universe Studies (SALTUS)* which, if selected, would launch in 2032 for a 5-year baseline mission duration in a Sun-Earth halo L2 orbit. *SALTUS* will study a wide range of astrophysical objects, such as the youngest galaxies, protoplanetary disks, and a variety of solar system objects including planets, moons, Kuiper Belt Objects, comets, and more. These studies will probe formation and evolution to answer fundamental questions about our cosmic origins, addressing key science aligned with the 2020 Astrophysics Decadal Survey[1] and the 2013 Astrophysics Roadmap[2]. The following papers expand on the importance of *SALTUS* to NASA science objectives and provide detailed discussion of the *SALTUS* goals: Chin et al., "Single Aperture Large Telescope for Universe Studies (SALTUS): Probe Mission and Science Overview," *J. Astron. Telesc. Instrum. Syst.* (this issue); Anderson et al., "Solar System Science with the Single Aperture Large Telescope for Universe Studies (SALTUS) Space Observatory," *J. Astron. Telesc. Instrum. Syst.* (this issue); Walker et al., "The High Resolution Receiver (HiRX) for the Single Aperture Large Telescope for Universe Studies (SALTUS)," *J. Astron. Telesc. Instrum. Syst.* (this issue); and Roelfsema et al., "The SAFARI-Lite Imaging Spectrometer for the SALTUS Space Observatory," *J. Astron. Telesc. Instrum. Syst.* (this issue).

For the *SALTUS* observatory flight system, Northrop Grumman draws from its significant NASA space mission heritage and has selected the Northrop Grumman LEOStar-3 product line as the primary spacecraft architecture coupled with a scaled GEOStar-3 primary structure[3] to host the *SALTUS* Payload. We define the 'observatory' as the spacecraft and Payload architectures. The Payload consists of a suite of far-infrared spectroscopic instruments, the Cold Corrector Module (CCM), the inflation control system (ICS), the Sunshield Module (SM), the Warm Instrument Electronics Module (WIM), and the Primary Reflector Module (PRM). The primary instrument suite is comprised of, 1) the High Resolution Receiver (HiRX): a four-band, high sensitivity, high spectral resolution heterodyne receiver, and 2) SAFARI-*Lite*: a four-band, high sensitivity, moderate resolution grading spectrometer. The PRM is made up of a deployable 14-m off-axis parabolic reflector by way of an inflatable membrane (M1), a deployable, segmented boom, and a truss. The SM is a truss-based cylinder with two deployable sunshield layers (stowed in a hub), where layer 1 is ~48.5 m × ~19.2 m (~931 m$^2$) and layer 2 is ~50 m × ~20 m (~1,000 m$^2$) and facilitates the *SALTUS* cryogenic <45 K environment for the PRM, CCM, HiRX, and SAFARI-*Lite*. Specialized truss-based composite structures, including an instrument deck and Core Interface Module (CIM), have been designed specifically to facilitate a requirements-derived mechanical and thermal interface between the Payload and the spacecraft. Northrop Grumman will leverage extensive heritage and experience from L'Garde, who will provide the 14-m M1, and from NeXolve and Redwire Space, who will provide the SM. In addition to *SALTUS* systems engineering, Northrop Grumman will provide mission management, safety and mission assurance (SMA), and additional resources to the Mission Operations Center (MOC; operated by Northrop Grumman at the PI-institution University of Arizona's Applied Research Building).

While the Northrop Grumman spacecraft architecture is more than capable of meeting mission requirements (e.g., orbit, environment and duration, launch vehicle (LV) compatibility, communications, etc.), *SALTUS* instrument-unique requirements have driven several key technical challenges associated with the observatory architecture, namely, i) pointing stability (to allow the CCM to capture light from M1 and correct for wavefront error for diffraction limited performance), ii) thermal stability and control (to facilitate a stable M1 temperature <45 K; a ΔT~250 K over ~3 m between the "warm side" and "cryogenic side" of the observatory, separated by the sunshield), and iii) lifetime, driven by a two-consumable system (first, sufficient propellant – for orbit transfer, station keeping, desaturation, and slew/settle times for observational efficiency; and second, sufficient inflatant gas (helium) – to maintain M1 inflation pressures that meet optical imaging performance). When coupled with the power of the corrective optics in the CCM and the performance offered by the Northrop Grumman spacecraft bus, the observatory architecture provides a robust solution for both the Guaranteed Time Observation (GTO; ~30%) and the Guest Observer (GO; ~70%) Astrophysics Probe Explorer (APEX) mandated observing programs.

The outline of this paper is as follows. In Sec. 2, we describe the *SALTUS* mission implementation, including general requirements, mission traceability, launch, and orbit. Section 3 discusses the observatory design and the flight system. In Sec. 4, we expand in detail on the sunshield module design. Finally, Sec. 5 details the *SALTUS* lifetime, constrained by the propellant and inflation control systems, and discusses observational efficiency, momentum unloading, and the effects of micrometeoroids (MMs) on M1.

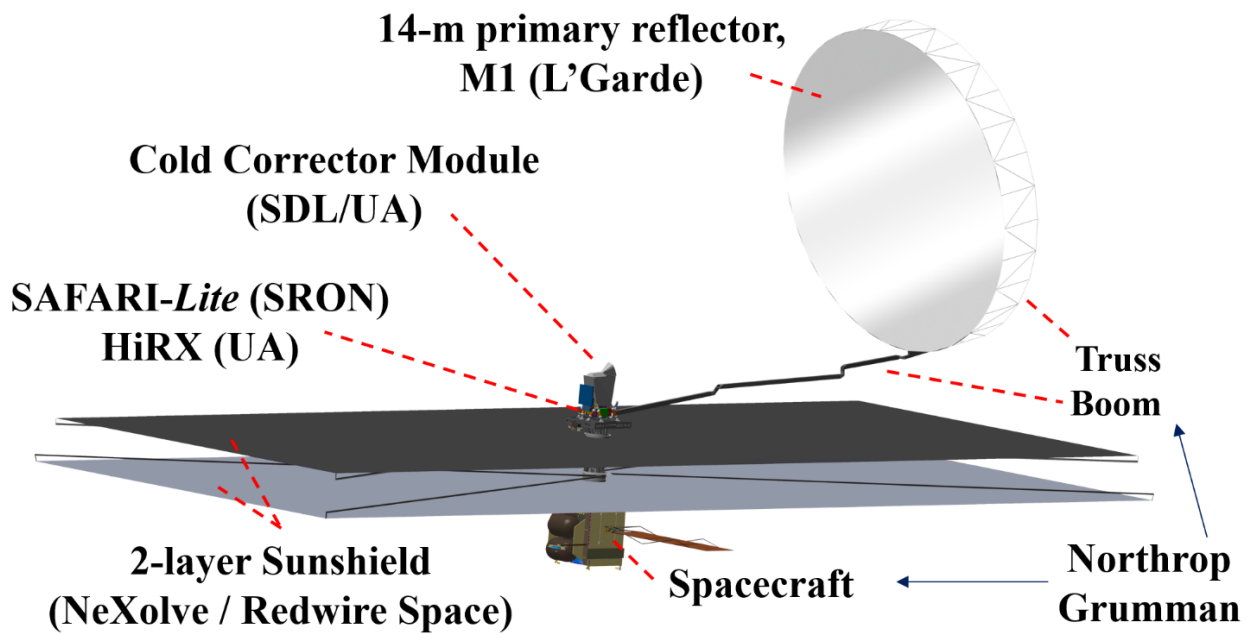

**Fig. 1** The *SALTUS* observatory. The partners listed are: University of Arizona (UA; PI-institution and responsible for HiRX), Space Dynamics Laboratory (SDL; Payload/CCM manager), SRON Netherlands Institute for Space Research (SAFARI-*Lite*), L'Garde (M1), NeXolve and Redwire Space (Sunshield Module). Northrop Grumman provides mission management, SMA, and resources to the MOC, and is responsible for the spacecraft and observatory architecture at large, including the composite instrument deck, CIM, ICS, deployable boom, and truss.

## 2 Mission Implementation

### 2.1 Concept Description and Mission Traceability

**Table 1** Summary overview of *SALTUS* mission requirements.

| Mission Design Requirements | Spacecraft Requirements | Ground System Requirements | Operations Requirements |
|---|---|---|---|
| - Orbit: Sun-Earth L2 halo<br><br>- 5-year mission duration<br><br>- Compatible with LV per AO guidance<br><br>- Launch NLT July, 2032 | - Propellant & inflatant to support 5-year mission *(Sec. 5)*<br><br>- Perform station keeping & desaturation to maintain observing efficiency *(Sec. 5)*<br><br>- Pointing knowledge: 3.3 arcsec, 1σ<br>- Pointing accuracy: 10 arcsec, 1σ<br>- Pointing stability: 0.66 arcsec, 3σ (over 10 min) *(Sec. 3.2.7)*<br><br>- Temperature for Payload systems: <45 K (cryo side) *(Sec. 3.2.8)*<br><br>- Closed loop control of inflation to 5.1 Pa ± 5.1 mPa *(Sec. 5)*<br><br>- LEGS compatibility *(Sec. 3.2.6)* | - Compatible with LEGS ranging *(Sec. 3.2.6)*<br><br>- Archive science and housekeeping data *(Sec. 3.2.4)* | - Transfer to L2<br><br>- Plan station-keeping maneuvers as needed<br><br>- Plan timing of science targeting, comm events, desaturation, calibrations, etc.<br><br>- Predict wheel momentum growth and plan desaturation maneuvers<br><br>- Sun remains at 90 degrees +/- 20 degrees of the observatory boresight and in the -Z direction |

With a launch readiness date (LRD) no later than July 2032, *SALTUS* launches from Kennedy Space Center (KSC) with a launch energy (C3) of -0.6 km$^2$/s$^2$. Launch is followed by a two-month commissioning and orbit transfer phase that includes multiple delta-v (ΔV) burns to achieve a trajectory toward the destination Sun-Earth L2 halo orbit. Following HiRX and SAFARI-*Lite* instrument commissioning, *SALTUS* begins its 5-year baseline primary mission. Figure 2 (a) illustrates the orbit's maximum 1.8 million km slant range from Earth with a maximum 25º Sun-spacecraft-Earth angle chosen to simplify the orbit design thus reducing the required orbit transfer ΔV. Earth eclipses are avoided with minimal ΔV, and the halo orbit is sized as such. The spacecraft attitude keeps the Sun at ~90º off the observatory boresight and limits rotation to ±20º of pitch and ±5º of roll – this is critical for maintaining the <45 K cryogenic environment on the M1 side of the sunshield and this orientation is maintained throughout the entire mission lifetime. Navigation requirements are easily met via Lunar Exploration Ground Sites (LEGS) ranging, where data volume and required data rates result in a communications system that fits within bandwidth limitations, as discussed in Sec. 3. Operations require a weekly total ground contact of 7h where the science team generates the target list well in advance of the contact. This can be adjusted in Phase A if necessary. The mission operations concept architecture is shown in Fig 2 (b).

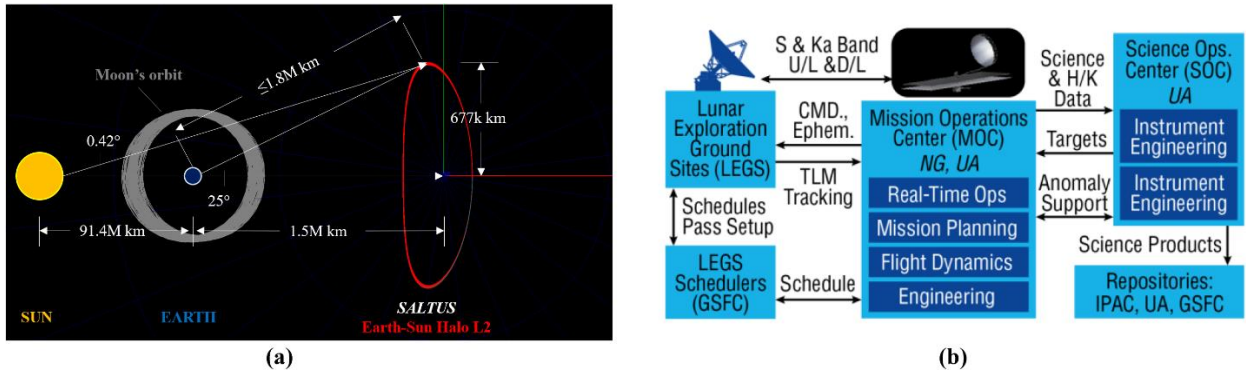

**Fig. 2 (a)** *SALTUS* Sun-Earth halo L2 orbit. The orbit has been sized such that Earth eclipses are avoided with minimal ΔV. The Moon's orbit is shown for scale. **(b)** *SALTUS* mission operations architecture, jointly supported by Northrop Grumman and University of Arizona's Mission Operations Center (MOC), Science Operations Center (SOC), and the Lunar Exploration Ground Sites (LEGS).

The *SALTUS* instantaneous field of regard (iFOR) is a circumpolar band of 40° width in the ecliptic frame, resulting in two continuous viewing zones of 20° around the ecliptic poles and full sky coverage in six months. This coverage is possible based on the attitude capabilities outlined above, coupled with the ability of the spacecraft to rotate 360° around the Sun line. The propulsion system performs momentum desaturation and station keeping maneuvers periodically throughout the mission, as described in Sec. 5. The overall *SALTUS* pointing approach combines the spacecraft attitude determination and control system (ADCS) and the fast steering mirror on the CCM. The spacecraft is required to point the CCM within 60 arcsec of its intended target, which is well within the scan range of the CCM's fast-steering mirror. Generally, targets outside the solar system are effectively inertially fixed and are treated as such for ADCS sizing. The process is repeated via stored command for moving solar system targets. Once a target is acquired, the observatory must have <0.5 arcsec pointing stability over a 20-second period in between fast steering mirror adjustments. The pointing stability requirement is an allocation of the overall transfer efficiency budget for the CCM. Note, the CCM transfer efficiency accounts for factors that ultimately determine the photon flux at each of the detectors – see Kim et al., "14-m aperture deployable off-axis far-IR space telescope design for SALTUS observatory." *J. Astron. Telesc. Instrum. Syst.* (this issue) for CCM capabilities and optical performance.

Since the availability of targets varies seasonally, with some periods of the mission more target-rich than others, the observation time requirement flows down to a conservative allocation of 60% operational target efficiency as discussed in Sec. 5 and in Chin et al., "Single Aperture Large Telescope for Universe Studies (SALTUS): Probe Mission and Science Overview," *J. Astron. Telesc. Instrum. Syst.* (this issue) . Table 1 shows as summary version *SALTUS* mission traceability matrix (MTM), which includes mission design requirements, spacecraft requirements, ground system requirements, and operations requirements. We have included references in parenthesis to sections in this paper where spacecraft and ground system requirements are addressed in detail.

## 2.2 Launch, Orbit, and Basic Operations

*SALTUS* will launch on a "Performance Upper (Std. PLF)" NASA-provided LV and will perform orbit transfer to the Sun-Earth L2 Lagrange point. This orbit was chosen for a variety of reasons, including no Earth/Moon eclipses, good stray light conditions, good thermal stability conditions, minimal ΔV station keeping, as well as the potential for enabling scientific synergy with the James Webb Space Telescope (JWST) and other missions. We conducted launch window analysis to offer a wide number of launch opportunities based on determining an optimal transfer trajectory design from an Earth departure asymptote to the desired halo orbit using GMAT – General Mission

Analysis Tool[4]. We considered a launch window every day for one year from the LRD of 01 JUL 2032 through 30 JUN 2033. Launch inclinations not accessible from KSC were discarded. Additionally, we removed cases where the required maneuvers (including mid-course correction (MCC) plus halo insertion) have a $\Delta V > 55$ m/s or a C3 > -0.6 km$^2$/s$^2$. Our analysis yielded 104 viable launch windows that met these criteria between 07 JUL 2032 and 29 DEC 2032. This $\Delta V$ budget is shown in Table 2 reflecting the 5-year baseline mission duration.

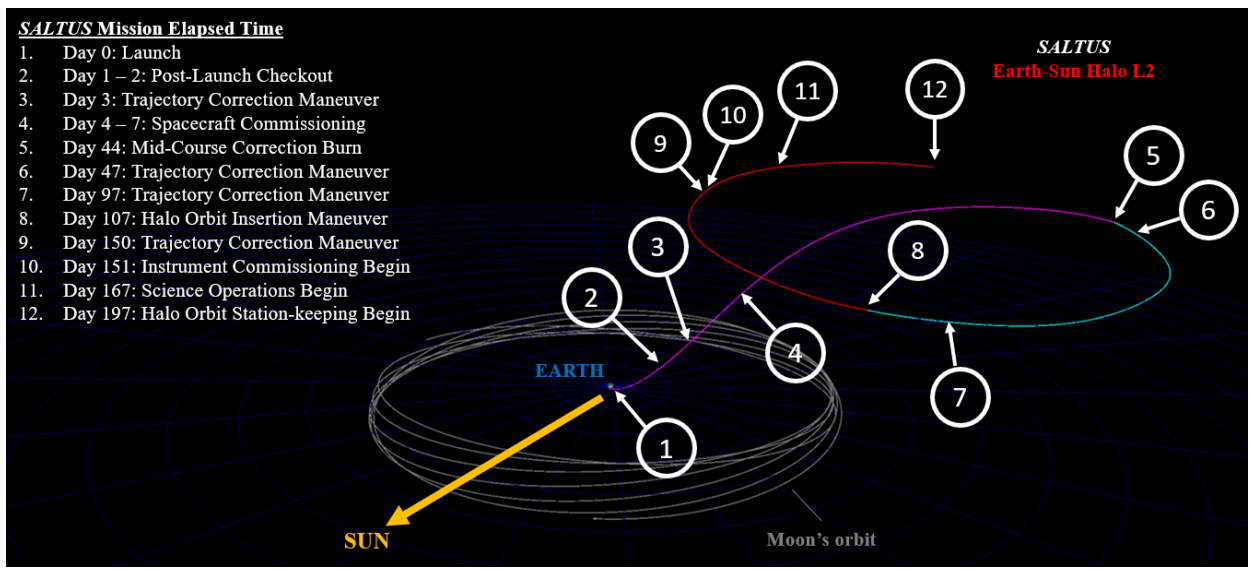

**Fig. 3** *SALTUS* launch into a Sun-Earth L2 halo orbit annotaed with maneuvers and sequence of activities to prepare for science operations. The Moon's orbit is shown for scale.

Following launch (L+0 days) and LV separation, the observatory autonomously deploys its solar array and enters a stable Sun-pointed mode. The operations team perform an initial checkout in preparation for a LV trajectory correction burn within the first three days after launch. The orbit transfer consists of the LV trajectory burn, a MCC maneuver ~L+44 days (resulting in a coincident trajectory with a stable low-energy transfer), and finally an injection burn L+107 days to place the observatory in the destination Sun-Earth halo L2 orbit. Trajectory Correction Maneuvers (TCM) are performed as needed during the transfer with LEGS ranging providing requisite orbit knowledge to support the transfer. The start of instrument commissioning does not commence until ~L+151 days, and is immediately followed by science operations, baselined at ~L+167 days. The orbit meets its Sun-spacecraft-Earth angle and maximum range requirements at 25º and 1.8 million km, respectively. Figure 3 depicts the orbit transfer and mission timeline graphically with appropriate labeling.

Instrument commissioning is executed after several observatory subsystem deployments – primarily, PRM and SM deployments after the CCM has sufficiently cooled to allow cryocooler operation. The PRM and SM are not deployed until all TCMs have been completed – this avoids excessive torque on the observatory during maneuvers, corrections, or insertion, where the observatory only undergoes station keeping after full deployment. Nominal operations consist of repeated activities, as follows: i) observing science targets (a typical observation of ~5h each), ii) momentum desaturation, iii) and slewing (including settling time) to the next target based on concept of operations (CONOP) for GTO and/or GO programs. M1 is under active inflation control during an observation to meet optical imaging performance, where inflation pressures are increased to 5.1 Pa ± 5.1 mPa to meet this performance just prior to observation onset. A detailed description of the inflation scheme, designed to maximize/extend *SALTUS* lifetime by only consuming inflatant gas during science data collection is outlined in Arenberg et al., "Design, Implementation, and Performance of the Primary Reflector for SALTUS," *J. Astron. Telesc. Instrum. Syst.* (this issue). Throughout a week of operations, the observatory slews to point its high gain antenna (HGA) toward the Earth to perform ground contact (total 7h per week, appropriately distributed via science CONOP) of high-rate Ka-Band data downlink and command uplink followed by low-rate S-band (low-bandwidth) LEGS ranging. An additional 7h per week of S-Band coverage has been budgeted for state-of-health (SOH) monitoring and will be optimized for efficiency and operational safety. Minor station keeping is scheduled every ~90 days. We discuss the operational on-sky efficiency of *SALTUS* in Sec. 5.

Finally, we impose two relevant mission requirements for navigation operations. First, the observatory must avoid drifting away from L2, and second the observatory must maintain communications. The link described above is sized to support simultaneous command, telemetry, and ranging, and we expect to perform LEGS ranging to these navigation-driven requirements. We also considered SM-induced orbital perturbation by assessing solar radiation pressure (SRP). This analysis was carried out in GMAT where the model included a simple 1,000 m$^2$ SRP plate perturbation. We found the effect of SM SRP is <2 m/s over the entire mission duration and is captured in our $\Delta$V calculations in Table 2. We also included $\Delta$V from momentum unloading (MU) which we found to be ~0.03 m/s per day. Station keeping (for 5 years) include station keeping maneuvers to maintain the halo orbit and the above corrections due to MUs. From the frequency of ranging contacts (at least two per week), and the minor SM SRP perturbations and MU described here, the $\Delta$V requirements are shown to be modest. Therefore, the orbit drift is small and a large impact to navigation performance is not expected. A detailed range accuracy requirements study will be completed in Phase A.

**Table 2** *SALTUS* $\Delta$V budget.

| Event | $\Delta$V |
|---|---|
| Trajectory Correction Maneuver (TCM) 1 | 25.0 m/s |
| Mid-Course Correction (MCC) | 66.4 m/s |
| TCMs 2-4 | 50.0 m/s |
| Halo Insertion | 17.1 m/s |
| Station Keeping (for 5 years) | 14.8 m/s |
| **Total** | **173.4 m/s** |
| Contingency | 30% |
| **Total (incl. Contingency)** | **225.4 m/s** |

# 3    Flight System Design

## 3.1    Overview

The *SALTUS* flight system is designed to meet the mission requirements outlined in Table 1 with margin while maximizing heritage and operational flexibility. The flight system is based on the Northrop Grumman LEOStar-3 product line, a platform with extensive heritage supporting successful NASA Category 2 programs both in Earth Science (e.g., JPSS-2, Landsat 9, ICESat-2) and Astrophysics (e.g., FERMI, Swift). Current Northrop Grumman programs continue to develop and improve the platform. Due to observatory loads and volume, *SALTUS* packages the LEOStar-3 architecture within a scaled primary structure based on the GEOStar-3. Payload interfaces are well defined and accommodated with standard protocols. In this section, we expand on the spacecraft, the Payload and SM interfaces, and all key flight system subsystems. We also discuss the PRM since Northrop Grumman is responsible for PRM deployables. The ICS is described in Sec. 5 since it directly relates to mission lifetime which Sec. 5 focuses on. We show a summary flight system block diagram in Fig. 4, the spacecraft stowed and deployed configurations in Fig. 5, summary technical budgets in Table 3 (mass), Table 4 (data), Table 5 (power), and Table 6 (link), and a flight system margin summary in Table 7.

## 3.2 Spacecraft Bus & Subsystems

### 3.2.1 Structures

The GEOStar-3 structure, and mechanical subsystem support the *SALTUS* Payload, provide the primary load path between the bus and LV, and the means of attachment for four required helium inflatant tanks for the ICS. The structure also supports the propulsion subsystem and provides ample mounting locations and radiative surfaces for flight system and Payload components. At the aft end, a heritage launch adaptor facilitates the interface between the spacecraft and LV.

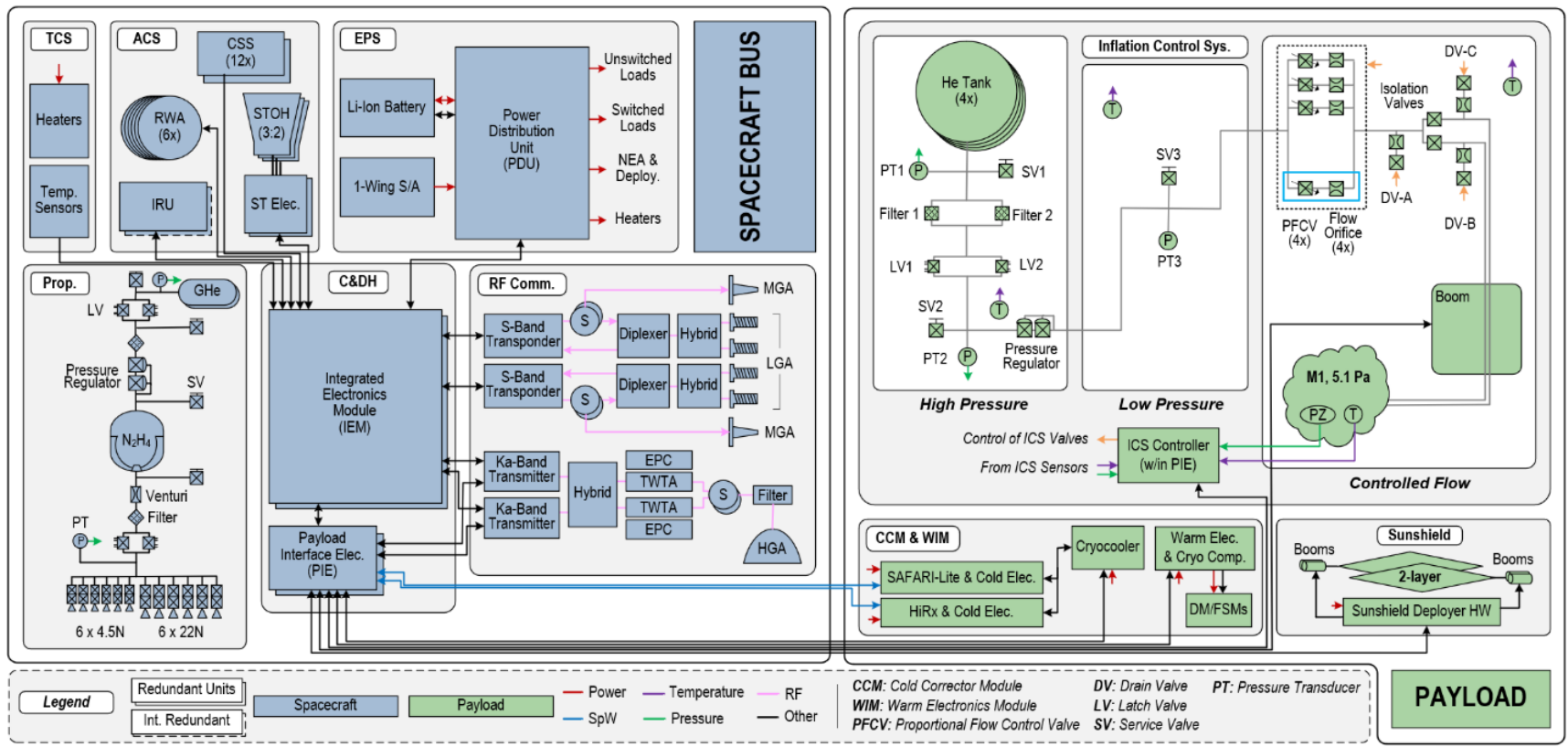

**Fig. 4** *SALTUS* flight system block diagram. The legend indicates selectively redundant or internally redundant components. LEOStar-3 avionics (IEM) are coupled with a Payload Interface Electronics (PIE) module to meet *SALTUS* science and mission requirements. Both the inflation control system (ICS) and Sunshield Module (SM) are captured under the Payload, as well as the Cold Corrector Module (CCM), Warm Instrument Electronics Module (WIM), and Primary Reflector Module (PRM). The ICS includes four tanks (helium gas) mounted externally to the GEOStar-3 primary structure. The CCM resides in the cryogenic environment (M1 side, <45 K), whereas the WIM resides inside the bus structure in the warm environment for thermal considerations (spacecraft side, ~310 K).

*3.2.2 Spacecraft to Payload Interface*

A fully composite truss-based structure – the CIM – provides the means of mechanical attachment and thermal separation between the Payload and spacecraft, as shown in Fig. 5 (c). This truss structure is based on JWST heritage and is tuned to be near zero coefficient of thermal expansion (CTE) in the cryogenic environment (see Sec. 3.2.8 regarding thermal stability). The CIM not only provides launch load capabilities and on-orbit thermal stability, but it also includes accessibility during I&T activities to route and thermally isolates the multiple cables and lines that span the spacecraft and Payload. On top of the CIM, we have designed a truss-based instrument deck leveraging similar JWST heritage and materials. The instrument deck provides the primary mechanical means of attachment for the CCM, ACS bench, and deployable PRM. Like the CIM, it is also tuned to near zero CTE at cryogenic temperatures to ensure thermal stability and to minimize pointing errors to meet pointing requirements. The SM is also a part of the *SALTUS* Payload and was designed to interface directly with the CIM and spacecraft maintaining an ideal load path with high stiffness and thermal performance.

**Table 3** *SALTUS* mass budget summary. (1) Current Best Estimate. (2) Maximum Expected Value.

| Mass Element | CBE (kg) (1) | Contgy. (%) | MEV (kg) (2) |
|---|---|---|---|
| Spacecraft | 2090.0 | 10.2% | 2304.0 |
| Payload | 939.8 | 20.8% | 1135.6 |
| **Total Observatory Dry Mass** | **3029.8** | **13.5%** | **3439.6** |
| LV Performance to C3 = -0.6 kg²/s² | | | 6846.0 |
| Helium Inflatant | | | 200.0 |
| Hydrazine Propellant | | | 1093.7 |
| Pressurant | | | 5.8 |
| **Total Observatory Wet Mass** | | | **4739.1** |

*3.2.3 Propulsion*

The Propulsion subsystem performs all maneuvers to reach the Sun-Earth halo L2 orbit as well as MUs and orbit maintenance throughout the mission. The subsystem is a mono-propellant hydrazine system sized to 1093.7 kg maximum expected value (MEV) with 5.8 kg MEV of pressurant based on flight-proven components. Six 22 N ΔV thrusters perform all impulsive maneuvers as well as X/Y attitude control during these maneuvers. Six 5 N ACS thrusters operate for Z-axis control during impulsive maneuvers and for MUs. All thrusters mount on the aft end of the spacecraft bus with the ΔV thrusters oriented to provide thrust along the +Z axis and ACS thrusters oriented to provide attitude control in all three axes. The propellant budget utilizes the MEV launch mass, 3σ low thruster performance, and maturity-based contingencies as well as efficiency factors such as misalignments, flight dynamics errors (relating to ΔV requirement uncertainty), and residual fuel.

*3.2.4 Command & Data Handling*

The Command & Data Handling (C&DH) subsystem provides command, control, and data handling capabilities within the *SALTUS* flight system via block-redundant Integrated Electronics Modules (IEMs) and block-redundant Payload Interface Electronics (PIEs). The IEM handles all core command, control, and data handling capabilities and provides triple modular redundancy (TMR) storage for the spacecraft and Payload SOH data storage, providing ample data storage margin. The IEM is augmented with block-redundant PIE units that provide command and telemetry interfaces for each element of the Payload and to the redundant Ka-Band System for downlink of stored SOH and science data. The PIE also provides the ICS Controller functionality, monitoring analog telemetry from the ICS and maintaining pressure within M1. Additionally, the PIE contains sufficient flash storage with TMR capability for science data storage. This storage capability provides ample margin for science data, as shown in Table 4. The *SALTUS* PIE is derived from similar units flown on Landsat 9 and JPSS-2 that provided mission-unique functionalities to their respective Payloads and redundant interfaces to the IEM.

**Table 4** *SALTUS* data budget summary.

| Parameter | Value |
|---|---|
| **Data Rates** | |
| SAFARI-*Lite* Science Data Rate (kbps) | 300.0 |
| HiRX Science Data Rate (kbps) | 250.0 |
| Bounding Science Data Rate (kbps) | 300.0 |
| Spacecraft SOH Data Rate (kbps) | 7.7 |
| Instrument HK Data Rate (kbps) | 0.3 |
| Total SOH Data Rate (kbps) | 8.0 |
| **Data Storage** | |
| Time between Data Downlinks (days) | 7.0 |
| Science Data to Store (Gbit) | 181.6 |
| Science Data Storage Available in PIE (Gbit) | 384.0 |
| Science Data Storage Margin (%) | 112.0 |
| SOH Data to Store (Gbit) | 4.8 |
| SOH Data Storage Available in IEM (Gbit) | 21.0 |
| SOH Data Store Margin (%) | 334.0 |
| **Data Downlink** | |
| St. Data Vol. to Downlink, Sci + SOH (Gbit) | 186.3 |
| Downlink Information Rate (kbps) | 10,000 |
| Allocated Time to Downlink (hr) | 7.0 |
| Downlink Volume Capacity (Gbit) | 252.0 |
| Downlink Volume Margin (%) | 35.0 |

*3.2.5 Power*

The Electrical Power Subsystem (EPS) supplies power to the spacecraft and Payload, and consists of a single deployable, non-articulating solar array wing with a Power Distribution Control, and Charging Unit (PDU), and a Li-Ion Battery. Figure 4 expands on this architecture, Table 5 summarizes the power budget for driving steady-state modes, and Table 7 includes solar array and battery performance. The battery is sized to support $\Delta V$ maneuvers within the allowed depth-of-discharge (DOD), accounting for all battery usage efficiency factors as well as capacity fading at EOL. The battery provides greater margin to all other battery-powered modes. The solar array supports all mission phases and modes with margin, and solar cell efficiency losses for the expected 100º C operating temperatures have been captured. Sized for a maximum Sun angle of 25º off solar array surface normal (per Fig. 2), the non-articulating solar array permits pointing across the entire *SALTUS* FOR. The driving steady-state power mode is a SAFARI-*Lite* science observation with HiRX in stand-by with a simultaneous Ka-Band science data downlink. The solar array provides greater margin to all other steady-state mission modes. As a conservative approach, MU, and slew modes are on battery power because the solar array is not gimbaled but could both be powered by the solar array, if necessary, as noted in Table 5.

**Table 5** *SALTUS* power budget summary. The driving steady-state power mode is a SAFARI-*Lite* Science observation with HiRX in stand-by with a simultaneous Ka-Band science data downlink (DL). **(1)** With a maximum MU duration of ~4 minutes and considering MU and slew battery recharge. This sizing case assumes a nominal observation and a continuous 5h science data DL. **(2)** The nominal DL cadence of 7h per week; Science + DL is used as a conservative case. Assuming return to an observation without DL, the solar array has additional margin when considering battery recharge demand and correcting MU duration. **(3)** We assume MU and slew modes draw power from the battery to avoid sizing the solar array to infrequent modes (4x per day). These modes could both be powered by the solar array.

| Subsystem | Science + DL (HiRX) | | | Science + DL (SAFARI-*Lite*) | | | Orbit Average Power | | |
|---|---|---|---|---|---|---|---|---|---|
| | CBE (W) | Contgy. (%) | MEV (W) | CBE (W) | Contgy. (%) | MEV (W) | CBE (W) | Contgy. (%) | MEV (W) |
| Bus Total | 921.5 | 9.6% | 1010.4 | 925.0 | 9.8% | 1015.6 | 917.6 | 9.5% | 1005.0 |
| Payload Total | 730.4 | 19.5% | 872.9 | 845.9 | 23.5% | 1045.1 | 732.5 | 19.8% | 877.5 |
| Observatory Total | 1651.9 | 14.0% | 1883.3 | 1770.9 | 16.4% | 2060.6 | 1650.1 | 14.1% | 1882.5 |
| **Observatory Total + 30% Margin** | **2147.5** | **14.0%** | **2448.3** | **2302.1** | **16.4%** | **2678.8** | **2145.1** | **14.1%** | **2447.2** |

| **Solar Array Sizing Mode Considerations** | **Contgy. (%)** |
|---|---|
| Solar Array Margin at EOL for Array Sizing Case with Downlink **(1)** | 20.0% |
| Solar Array Margin at EOL for Array Sizing Case Without Downlink **(2)** | 39.0% |
| Solar Array Margin at EOL during Slew **(3)** | 8.0% |
| Solar Array Margin at EOL during Momentum Unload **(3)** | 38.0% |

*3.2.6 Telecommunications*

The Telecommunications subsystem provides both low-rate Tracking, Telemetry, and Commanding (TT&C) capabilities throughout all *SALTUS* mission phases and high-rate science data downlink capabilities, and as a result requires two distinct systems: a low-rate S-Band system and a high-rate Ka-Band system as shown in Table 6.

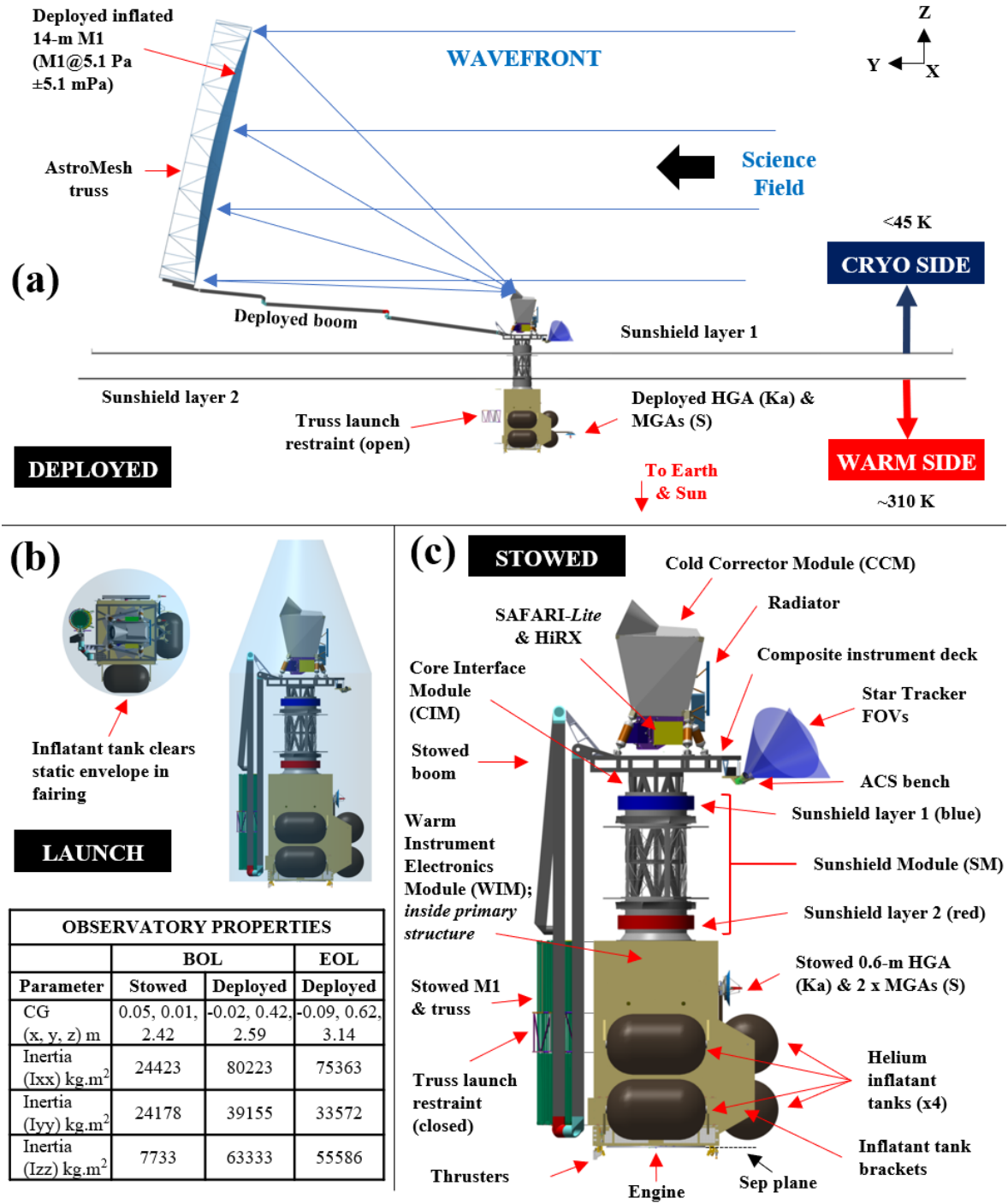

**Fig. 5 (a)** *SALTUS* deployed configuration. The "warm side" is Earth and Sun facing. M1 boresight is ~90º (±20º pitch, ±5º roll) to the Sun line, and the Z axis can rotate 360º around this line. **(b)** *SALTUS* launch configuration. The observatory fits in the static envelope of the LV fairing. We also show observatory Center of Gravity (CG) w.r.t. the spacecraft origin at the separation plane ("Sep plane"), and inertia in stowed and deployed configurations. **(c)** *SALTUS* stowed configuration. Sunshield layers are colored to indicate layer 1 (blue, cryo side) and layer 2 (red, warm side).

**Table 6** *SALTUS* link budget summary. (1) C + RS represents Convolutional + RS (255, 223). (2) Margin for TT&C link maintained at minimum 6 dB for ranging-induced losses. All other links maintained at minimum 3 dB.

| Parameter | Low-Rate / Safe | | TT&C | | Science |
|---|---|---|---|---|---|
| | Uplink | Downlink | Uplink | Downlink | Downlink |
| Frequency (MHz) | 2072.5 | 2287.5 | 2072.5 | 2287.5 | 26375.0 |
| Elevation Angle (°) | 5 | 5 | 5 | 5 | 5 |
| Maximum Range (km) | 1,900,000 | | | | |
| Information Rate (kbps) | 1.0 | 1.0 | 2.0 | 8.0 | 10000.0 |
| FEC (1) | None | C + RS | None | C + RS | C + RS |
| Bit Rate (kbps) | 2.0 | 2.3 | 2.0 | 18.3 | 22870.0 |
| Modulation | BPSK | BPSK | BPSK | BPSK | QPSK |
| Antenna Coverage | 80% | 80% | 50% | MGA | HGA |
| Spacecraft EIRP (dBm) | N/A | 38.3 | N/A | 52.8 | 88.0 |
| Spacecraft G/T (dB/K) | -33.9 | N/A | -30.8 | N/A | N/A |
| Ground Station | LEGS | | | | |
| BER | $1 \times 10^{-6}$ | | | | |
| Link Margin (dB) (2) | 6.8 | 3.2 | 6.9 | 8.6 | 5.0 |

The S-Band system is designed for compatibility with select ground and space assets during Launch and Early Orbit Phases (LEOPS) and the NASA Near Space Network LEGS network during cruise and science phases. A pair of low-gain, quadrifilar helix antennas (LGAs) mount onto opposite sides of the spacecraft bus providing near 4π-steradian coverage during uplink and downlink. Both S-Band receivers remain powered throughout all flight system modes, while the active S-Band transmitter is powered only during post-separation and rate capture, safe mode, and scheduled passes. Each S-Band channel is augmented with a medium-gain, monofilar antenna (MGA) that points in the same direction as the 0.6-m Ka-Band high gain antenna (HGA) to provide higher rate S-Band capability for TT&C during science data downlink.

The Ka-Band system is compatible with the LEGS network for science data downlink and draws heritage from the JPSS-2 mission. The PIE and Ka-Band transmitter modulate and condition the data stream for downlink to LEGS. The body-fixed HGA has a beamwidth greater than the angular size of Earth at the Sun-Earth halo L2 orbit, and points in the aft direction. The HGA and MGAs are stowed against the spacecraft bus during launch and are boom-deployed during the cruise phase. Throughout a period of one week (optimized via CONOP), *SALTUS* points its HGA toward the Earth to perform ground contact of 7h total high-rate Ka-Band data downlink and command uplink followed by low-bandwidth, low-rate S-band via LEGS ranging. As noted in Sec. 2.2, an additional 7h per week of S-Band coverage has been baselined for SOH monitoring and other communications as required by mission needs.

### 3.2.7 Attitude Control

The ACS provides pointing stability, control, and knowledge for science observation, while also providing sufficient agility across the FOR. *SALTUS* pointing requirements (as shown in Table 1) are readily met with margin on the LEOStar-3 platform as a result of carefully selected star trackers (STs), inertial reference units (IRUs), reaction wheel assembles (RWAs), and Northrop Grumman flight-proven algorithms (e.g., ICESat-2, Landsat 8/9). We show the ACS bench location and approximate ST FOVs in Fig. 5 and provide performance margins for each ACS mode in Table 7. Attitude stability and control is maintained and stabilized throughout all mission phases, beginning with LV separation. During $\Delta V$ maneuvers, the ACS utilizes its $\Delta V$ mode providing slew capability to and from the maneuver attitude. In spacecraft safe mode, the ACS nulls spacecraft rates, points the -Z axis at the Sun while keeping the solar array Sun pointed with a slow roll induced about the Z axis. Safe mode therefore maintains a power-positive, thermally safe, commandable attitude that avoids *SALTUS* pointing constraints. This effectively eliminates solar pressure induced momentum accumulation and keeps the spacecraft approximately Earth pointed as well.

Attitude sensor and CCM misalignments are minimized by collocating the dedicated, thermally controlled ACS bench with the CCM on the instrument deck and the ACS sensors are oriented to provide attitude knowledge in all three degrees of freedom (DoF). Six RWAs manage momentum and slew the observatory, and the six 5 N ACS thrusters provide MU as needed. SRP, driven by the sunshield area (see Sec. 4), dominates the environmental torques during science operations and remains nearly constant during an observation. Current predictions of environmental torques acting on the *SALTUS* observatory require a MU event every 5.4h – 8.0h (BOL – EOL). A typical science observation is estimated to be ~5h, and since contiguous observations are not a requirement, this MU cadence is acceptable. We discuss MU in greater detail in Sec. 5.

The LEOStar-3 platform has extensive heritage providing arcsecond-level precision-pointing and pointing stability capabilities, and readily meets the *SALTUS* pointing requirements with margin. Using Northrop Grumman's flight-validated simulation capabilities for attitude estimation and control errors and allocations based on heritage programs, the predicted performance of the flight system ACS meets *SALTUS* needs with robust margin, as shown in Table 7.

### 3.2.8 Thermal Control

**Overview**. The Thermal Control Subsystem (TCS) employs a passive, cold-biased design with radiative cooling and heaters to keep *SALTUS* flight system components within acceptable temperature limits during all operational mission modes. As noted previously, the +Z side of the observatory faces away from the Sun during science observations (in fact, once the cryogenic environment has been established, +Z will maintain this attitude until EOL ±margin in roll, pitch, and yaw), providing a thermally stable environment and ample exterior radiator area. Radiators and heaters are sized to provide large margin against qualification temperatures, and all heaters are sized with appropriate duty cycles under worst-case cold and low voltage conditions.

The *SALTUS* spacecraft is located on the Sun-side of the SM, as shown in Fig. 5 (a). As a result of large sunshield area, this configuration causes the external surfaces of the spacecraft to be subjected to high amounts of, (i) sunlight reflected off the deployed SM, and ii) IR-backloading (radiant heat) emitted from the deployed SM surfaces that are in-view of spacecraft surfaces. The magnitude of these impacts on spacecraft and SM thermal performance can vary significantly depending upon the size, configuration, and external coatings of the deployed SM, the configuration of the spacecraft, as well as the interfacing geometry and hardware configuration between the two. Consequently, early thermal, and mechanical design efforts have considered interface designs between the spacecraft and SM via the CIM (details discussed in Sec. 3.2.2) and bus component operating temperatures, such as batteries and solar cell efficiencies have also been captured in sizing.

Fortuitously, the JWST design team encountered similar interface considerations for its spacecraft and sunshield. In the JWST case, thermal design necessitated the inclusion of deployable radiator shields (DRS) and fixed radiator shields (or, non-deployable radiator shields, FRS). Specialized insulation, coating designs, and lightweight structures resulted in a highly efficient solution, not only in its deployed (in-flight) state, but also in stow. The DRS provided valuable functionality by covering radiators to reduce spacecraft heat rejection capability, thusly reducing heater power for the stowed configuration when less power is available (prior to solar array deployment). In Phase A, Northrop Grumman will leverage this significant domain knowledge from this JWST proven concept to guide an efficient, robust, and low risk *SALTUS* thermal interface design to address the warm side bus thermal design challenges in full.

**Thermal Environments**. The observatory has two distinct thermal zones: i) the warm side, on the -Z side of the sunshield, and ii) the cryogenic side, on the +Z side of the sunshield, as shown in Fig. 5 (a). Notably, for thermal concerns, the warm side contains the WIM and solar array, whereas the cryogenic side contains the CCM, PRM, and ACS bench. As discussed in Sec. 3.2.2, these environments are mechanically and electrically connected via the CIM. Using the same analytical tools as JWST, we predict an equilibrium temperature of ~310 K on the warm side, where radiative cooling to deep space allows all components on the cryogenic side to cool to ≤45 K. This is critical for the exceptional IR sensitivity for *SALTUS* science. To fully understand the stepdown in temperature from ~310 K to ≤45 K, we constructed a cryogenic heat map of *SALTUS* as shown in Fig. 6. The HiRX and SAFARI-*Lite* focal planes require cooling to ~5.25 K, therefore we selected a build-to-print version of the JWST MIRI cryocooler to provide 31 mW of lift at 5.25 K, plus 110 mW of lift at 20 K for the SAFARI-*Lite* low noise amplifiers (LNAs). This performance includes >30% margin. We used Thermal Desktop modeling to ensure the CCM's design provided 500 mW of cooling at 45 K, and included two side-looking radiators, 1.5 m$^2$ and 0.5 m$^2$, respectively, to provide additional radiative cooling needs for HiRX. The MIRI compressor system resides on the warm side of the spacecraft (for thermal isolation) and is coupled to the WIM, where gas lines facilitate the pressurized helium to a valve at the 5.25 K stage of the system on the cryogenic side via the CIM. Northrop Grumman has extensive experience with the MIRI cryocooler and therefore understands power consumption well. Thus, a 2.7 m$^2$ radiator (coated with specialized paint) radiates to deep space to cool the WIM. Additionally, the instrument radiator 4 m$^2$ surface area provides 50% margin on these loads and will be placed orthogonally to the spacecraft-Sun axis for a clear view to deep space. All radiator sizing in this work is first order approximation, and will be refined and fully addressed in Phase A.

**ACS bench**: The ACS bench, a thermally controlled deck, contains three STs and an IRU as shown in Fig 5 (c). The bench is mounted on the underside of the composite instrument deck via titanium flexures for conductive isolation. Importantly, the ACS bench is completely out of the M1 and CCM cryogenic FOVs and draws heritage from an identical ICESat-2 design. The STs and IRU take advantage of two dedicated radiators allowing each to radiate heat to space – one sized to compensate for ~1 W per ST, and the other sized for the IRU's ~40 W at 20 C. Since the STs and IRU are used in all ACS modes, survival heaters maintain greater than qualification temperatures in event of anomaly. Additionally, ST and IRU blankets surround the units minimizing instrument deck radiative coupling. A detailed flexure/MLI design will occur in Phase A.

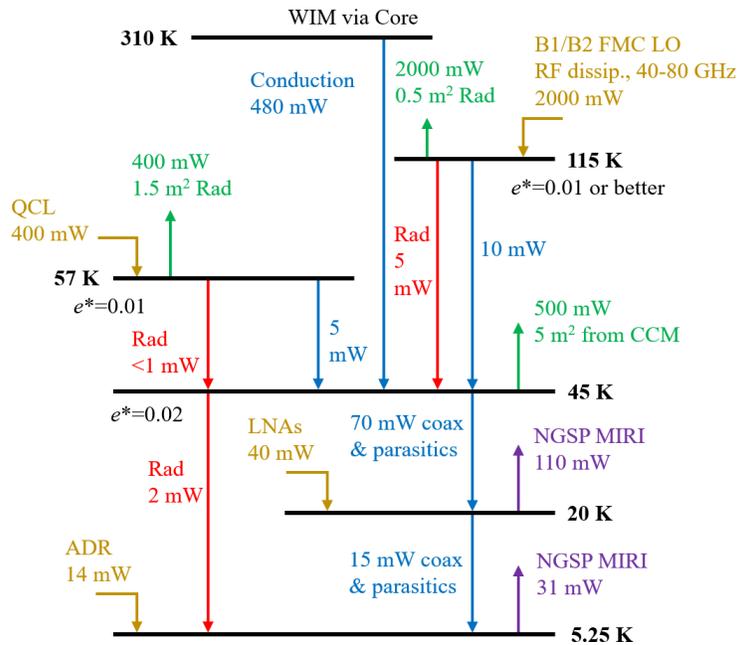

**Fig. 6** *SALTUS* cryogenic heat map. Acronyms in figure: ADR – adiabatic demagnetization refrigerator; CCM – cold corrector module; LNA – low noise amplifier; LO – local oscilator; QCL – quantum cascade lase; WIM – warm instrument module.

**<u>Sunshield Module</u>**: Since the SM allows direct radiative coupling between the warm side (spacecraft) and cryo side (Payload), other thermal control features are designed into the interface between the CIM and SM. Single and multiple layer radiative barriers (i.e., blankets placed near the top and bottom of the SM) are planned to eliminate radiative coupling with no impact to the design. Cable and harness traversing the warm to cryogenic environments will be flat and have high emittance wrapping to enhance thermal radiation. If needed a harness radiator will be added. For the harness in the CIM region, conductors, shielding, and insulation will be chosen to minimize conducted heat. An environmental shield limits radiative view factors of the warm CIM to the cryocooler refrigerant line, lowering radiative coupling and the possibility of icing.

The thermal effects of the SM on the spacecraft and solar array are also considered. We have used JWST flight actuals and found these data relevant due to similar sizing, space, and radiative coupling. We used the JWST peak observed operating temperature on the solar array as the nominal *SALTUS* case and sized the *SALTUS* single-wing solar array on that basis. Temperatures of thermally isolated hardware, like the boom, are driven by radiation which changes little over the *SALTUS* iFOR, see Arenberg et al., "Design, Implementation, and Performance of the Primary Reflector for SALTUS," *J. Astron. Telesc. Instrum. Syst.* (this issue). Therefore, material CTE matching in the composite boom was also considered. The boom has aluminum and titanium fittings which have a positive CTE. The composite's CTE is slightly negative and can be tuned to make the net boom CTE near zero over the relevant M1 temperature range. Even when the CTEs are tuned, the ability to perfectly match them is limited by the finite precision of CTE measurement, roughly 0.01 ppm/measurement. Since there are three materials, the minimum CTE would be $\sqrt{3}$ times this value, or ~0.02 ppm. For a ~17-m boom and the maximum temperature change of 1 K, say, the boom would change length less than 0.5 µm, which can be corrected by the optical system.

### 3.2.9 Flight Software and In-Flight Fault Management

Nearly all functionality required for *SALTUS* is already part of the baseline LEOStar-3 C&DH and ACS flight software (FSW). This FSW has a proven flight track-record from missions such as ICESat-2, Landsat 8/9, and JPSS-2, where *SALTUS* can leverage extensive re-use of existing FSW with modifications limited to *SALTUS*-unique functions. This setup offers modularity and flexibility for additional features/capabilities where reconfiguration is accessible on-orbit.

Redundancy via in-flight fault management is implemented throughout all subsystems, where standard single-point failure exemptions are acceptable for structure, fuel lines, tanks, and the HGA. Onboard fault detection via FSW as the primary, and hardware as the secondary detects and responds to anomalous conditions as identified. Flexible, heritage FSW TLM Monitoring (TMON) is used to manage onboard fault detection and correction with the list of critical TLM mission configurables. The FSW maintains a list of failure symptoms for each component. If TLM violates both the threshold and persistence check, the FSW TMON function responds by executing the appropriate command sequence. As noted previously, safe mode sheds all non-critical loads and achieves Sun pointing using the CSSs, and both S-Band receivers are always on. Returning *SALTUS* to science operations is performed by Mission Operations staff on the ground.

**Table 7** *SALTUS* Flight System margins summary. (1) LV performance capability to *SALTUS* orbit. (2) Includes full inflatant load of 200 kg. (3) Maximum dry mass assuming a full propellant tank.

| Parameter | Requirement | Performance | Margin |
|---|---|---|---|
| Observatory Wet Mass | 4739.1 kg | 6846.0 kg (1) | 45% |
| Observatory Dry Mass | 3439.6 kg (2) | 5204.1 kg (3) | 43% |
| Propellant Tank Capacity | 1093.7 kg | 1641.9 kg | 50% |
| Solar Array Performance (EOL) | 2211.0 W | 2874.9 W | 30% |
| Battery Capacity (BOL) | 96.0 A-hr | 134.0 A-hr | 40% |
| Pointing Accuracy [1σ] | 10.0" | 1.2" | 736% |
| Pointing Stability [1σ] | 2.0" | 0.66" | 189% |
| Pointing Knowledge [1σ] | 3.33" | 1.19" | 180% |
| Science Data Storage | 181.6 Gb | 384.0 Gb | 112% |
| Link Margin | TT&C Uplink | 0 dB | 6.9 dB | 6.9 dB |
| Link Margin | TT&C Downlink | 0 dB | 8.6 dB | 8.6 dB |
| Link Margin | Ka-Band Downlink | 0 dB | 5.0 dB | 5.0 dB |

*3.3 Primary Reflector Module (PRM)*

*3.3.1 Overview*

The *SALTUS* PRM is made up of three key components: the 14-m M1, the AstroMesh segmented boom, and the AstroMesh truss. The L'Garde M1 membrane is discussed in detail in Arenberg et al., "Design, Implementation, and Performance of the Primary Reflector for SALTUS," *J. Astron. Telesc. Instrum. Syst.* (this issue). This section will focus on the AstroMesh boom and truss, which Northrop Grumman is responsible for in this development.

The AstroMesh reflector was developed in the late 1990s[5] for the Thuraya constellation and has been refined over the last 30 years with over eleven successful mission deployments (see Table 8). The AstroMesh architecture is therefore well proven, inherently reliable, and offers robust deployment kinematics for a wide range of applications. The AstroMesh truss is designed as an efficient drum-like structure with high deployed stiffness, extremely low mass, and high thermal stability, making it a desirable option for the *SALTUS* use case. Indeed, its high structural stiffness allows for deployed performance measurements in 1G environments, further enhancing I&T capabilities. Crucially, the architecture was designed with scaling in mind, where an increase in truss outer diameter (aperture size) is implemented without significant change thus maintaining heritage and minimizing mission-unique development.

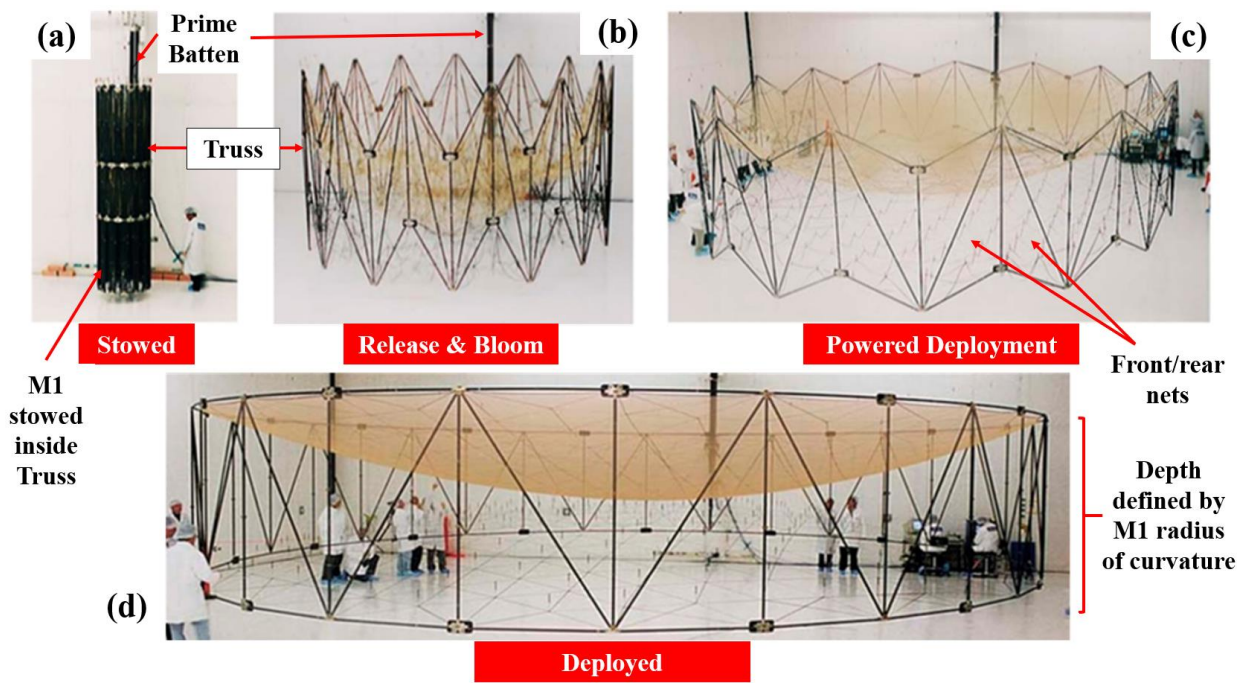

**Fig. 7 (a)-(d)** *SALTUS* AstroMesh architecture showing the stowed and deployed configurations, as well as two key steps in the truss deployment: release & bloom, and powered deployment. Note, the golden mesh reflective surface shown in this figure is illustrative only; the *SALTUS* M1 inflatable membrane will instead occupy this volume. Image credit: Northrop Grumman.

The basic AstroMesh architecture as shown in Fig. 7 consists of a flat perimeter truss that supports front and rear nets made of high-precision composite tension "webs", which define a geodesic surface. The depth of the truss structure is defined by reflector aperture and radius of curvature. Note, for the *SALTUS* application, these webs will be used for structure only and are not needed to form the surface geometry of M1 – this is driven solely by the ICS. Coupled to the truss are high-precision, high-heritage deployable booms, also scalable to meet increases in truss diameter without impact to heritage. The AstroMesh truss and boom combination is fully electrostatic damage (ESD) compliant with a demonstrated passive intermodulation (PIM)-free design. The gold-molybdenum mesh shown in Fig. 7 (b)-(d) is normally used for RF applications and is included in the figure for illustrative purposes only.

**Table 8** *SALTUS* AstroMesh truss and boom flight heritage. (1) The main change for *SALTUS* is a scaling of the truss for 12-14 m. (2) Demonstrated in aperture size range. The *SALTUS* boom will be scaled for *f/D*. (3) AstroMesh is Electrostatic Damage (ESD) compliant and verified Passive Intermodulation (PIM)-free by design. (4) Driven by aperture, not mission unique. (5) Includes 20% contingency.

| Parameter | SMAP | Inmarsat-4 | Inmarsat-6 | Alphasat | Thuraya | MB Sat | *SALTUS* |
|---|---|---|---|---|---|---|---|
| Aperture Size | 6 m | 9 m | 9 m | 11 m | 12 m | 12 m | 14 m (1) |
| Flown in Space | Yes | Yes | Yes | Yes | Yes | Yes | -- |
| **# Successful Deployments** | **1** | **3** | **2** | **1** | **3** | **1** | **--** |
| *f/D* (2) | 0.5-1.5 | | | | | | 1.24 |
| ESD/PIM (3) | Compliant: Verified by design and demonstrated | | | | | | Verif. by design |
| Aperture Deployment | Offset | | | | | | Offset |
| Reflector Mass (4) | 29 kg | 49 kg | 49 kg | 63.5 kg | 63.5 kg | 63.5 kg | 12 kg (5) |

*3.3.2 Heritage & TRL*

Table 8 highlights a range of space flight heritage missions that have successfully deployed the AstroMesh system. The *SALTUS* 14-m truss diameter is easily within the capabilities of the AstroMesh scalable design (as noted in Sec. 3.3.1) where all previous missions shown in Table 8 have demonstrated *f/D* ratios in line with the expected *SALTUS* M1 *f*-number. Previously demonstrated offset (off-axis) aperture deployments and reflector masses also introduce no additional development for *SALTUS* given the extreme lightweight mass of the M1 membrane. Table 8 also highlights a heritage assessment for the truss and boom with an associated TRL-6 assigned to each component. We expect to comply with all partial heritage elements, where the key required changes have been identified as, i) a scaling of the truss from demonstrated 12-m to 14-m geometry and adapted to *SALTUS* optical configuration (for radius of curvature), and ii) a scaling of the boom to the deployment length to meet the *SALTUS* M1 *f*-number, as well as spacecraft accommodation for inflatant lines associated with the ICS (see Sec. 5) and specific actuator step size implementation for boom alignment. These are considered low risk implementations and a standard part of mission-unique design for the AstroMesh system.

*3.3.3 Deployment Sequence*

Figure 8 illustrates the *SALTUS* PRM deployment sequence in six key steps. We note that each hinge will not be required to fully deploy (or fully precision-latch) before a subsequent hinge actuator motion – this flexibility provides an option later for positioning optimizations, clearance optimizations, assessment of moment of inertia impacts, etc. The finer details of the PRM deployment sequence will be refined in Phase A where this subsection presents the basic concept. Below we outlined the six-step sequential process through full PRM deployment:

- **STEP 1**: At launch, the *SALTUS* stowed truss containing the stowed M1 membrane is bound to the spacecraft at three attachment points, as shown in Fig. 8 (a). At the upper and lower end of the truss, a cable holds the assembly bundle into flexure-mounted "saddles" comprising upper and lower ridge node saddles. At the approximate center point of the bundle, a hinged "cradle", the Valley Node Cradle (VNC), restrains the truss, which is held closed with hold down and release mechanisms (HDRM's). Boom loads are shared by the root hinge, and four additional tie-down points.

- **STEP 2**: After launch (approximately L+1-2 days), ridge node cables and boom tie-down points release, thus minimizing thermal loads during subsequent maneuvers in flight to the Sun-Earth halo L2 orbit. Ridge node cables are reeled into spring-loaded mechanisms after they are cut. First motion occurs when the VNC releases. Kick-off springs on the Prime Batten propel the truss out of the VNC and rotates the truss around the boom elbow actuator which includes an integrated ratchet to prevent bounce-back. A stepper motor actuator deploys the first boom hinge. This sequence is shown in Fig. 8 (b).

- **STEPS 3-5**: Subsequent boom actuators are deployed in kind sequentially until the boom is fully deployed as shown in Fig. 8 (e). Each actuator deployment is choreographed to ensure there is minimal risk for interference with other spacecraft structures and to minimize disturbances to the spacecraft. This is typically a very slow and gentle process taking hours to days depending on the number of steps and verification pauses required, as well as the precision of the actuators.

- **STEP 6**: After the boom completes deployment STEPS 1-5, the truss deployment begins, starting with *Secondary Release*, as depicted by a real AstroMesh picture in Fig. 7 (b). This is the severing of a restraint cord about the center of the stowed M1 bundle allowing the M1 bundle to bloom. *Bloom* is the event coincident with and immediately following the *Secondary Release* as the truss expands rapidly and radially from the Prime Batten. This process is driven by strain energy in the stowed truss. After the dynamics of the *Bloom* have subsided, the truss deployment motors are activated and motorized deployment begins (Fig. 7 (c)). Dual redundant spoolers at the Prime Batten withdraws cable from either end of the single deployment cable that is routed through the truss diagonals. Both ends of the cable are pulled through cable tension sensing mechanisms for assessing progress and

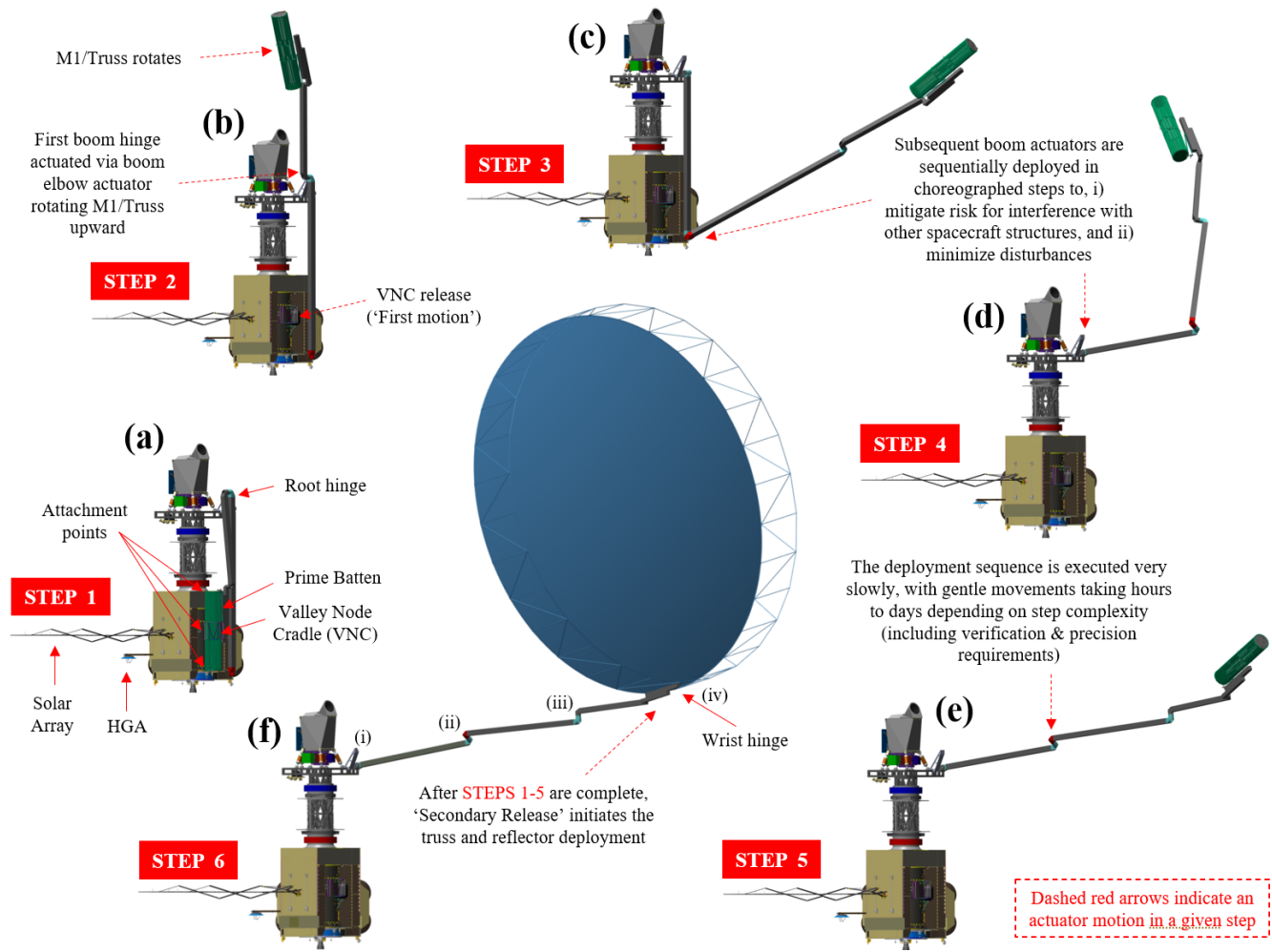

**Fig. 8 (a)-(f)** *SALTUS* Primary Reflector Module (PRM) deployement sequence. The solar array and HGA are deployed prior to PRM deployment. The sunshield is deployed following PRM deployment. This sequence is important to avoid a cryogenic PRM deployment (which is a more complex proposition). Note, each hinge will not be required to fully deploy or latch before the next hinge actuator in the sequence initiates its motion. This strategy allows for positioning optimizations, clearance optimizations, moment of inertia impacts, etc., and will be refined in Phase A. STEP 6 (f) (i-iv) shows the four actuator boom locations.

ultimately to provide the information required to terminate the deployment. As the cable is wound onto the spools, the diagonals become shorter, eventually latching into their final positions as the bays transition from partially deployed parallelograms to fully deployed rectangles. Figure 7 (d) shows a real laboratory demonstration of the truss fully deployed, and Fig. 8 (f) illustrates the fully deployed PRM with respect to the spacecraft which ends the deployment sequence. At this point, finer alignment and M1 positional adjustments with respect to the CCM begin.

Note, during deployment, the retention clips manage the PRM soft goods, which include the front and rear net as well as the uninflated M1 membrane. These release regularly and passively throughout the deployment. Once deployed and latched, the spoolers are backed off slightly to reduce cable tension and the truss deployment is completed.

*3.4 Observatory End-to-End Pointing Alignment and Stability*

Northrop Grumman conducted several studies that assessed stability provided by the ACS design, as well as stability of the bus structure, CIM, and PRM. In addition to addressing the *SALTUS* ACS system performance to meet mission requirements in pointing knowledge, accuracy, and stability (Table 7) which specifically addresses capabilities at the plane of the instrument deck, we must also consider *observatory end-to-end pointing alignment and stability* ultimately guided by the *SALTUS* alignment tolerancing process. This describes the maximum allowable magnitude of perturbation as a function of rigid body motion DoF to meet the *SALTUS* optical performance requirements. We refer the reader to the alignment tolerancing process as described by Kim et al., "14-m aperture deployable off-axis far-IR space telescope design for SALTUS observatory." *J. Astron. Telesc. Instrum. Syst.* (this issue). In this paper, we will address the mating of the spacecraft

ACS performance, including vibration and jitter, with the positional in-design accuracies of the AstroMesh boom, and the observatory settling time after a maneuver.

**Table 9** *SALTUS* observatory deployed structural mode estimates. The coupling of the low frequency first mode of the deployed sunshield layer with the low mass of each layer is not anticipated to have substantial effects on pointing performance. However, this will be included in a more detailed pointing vibration analysis in Phase A.

| Mode # | Frequency (Hz) | Description |
|---|---|---|
| 1 | 0.19 | M1 $\theta x$ rotation relative to spacecraft |
| 2 | 0.29 | M1 $\theta y$ rotation relative to spacecraft |
| 3 | 0.51 | M1 $\theta z$ rotation relative to spacecraft |
| 4 | 0.91 | M1 x translation relative to spacecraft & slosh |
| 5 | 0.91 | M1 y translation relative to spacecraft & slosh |
| 6 | 1.11 | M1 z translation relative to spacecraft & slosh |
| 7 | 1.11 | Propellant slosh |
| 8 | 3.13 | Solar array primary bending |
| 9 | 8.37 | Solar array secondary bending |
| 10 | 18.24 | Solar array rotational mode |
| 15 | 31.17 | First spacecraft bus structural mode |

To capture this trade at the observatory level, it is useful to think about the hierarchy as follows: i) spacecraft level (ACS, including pointing performance and vibration/jitter), ii) PRM level (root hinge, boom, M1/truss), and iii) CCM level (optical elements M2…M7, described below). The spacecraft provides attitude performance measured from the instrument deck under some vibrational load, and the PRM provides alignment with respect to the CCM measured from a combination of, i) performance from the root hinge on the instrument deck, and ii) the positioning accuracy and linear/rotational step size of the boom actuators. The superposition of these performances defines the position of the M1 boresight with respect to the M2 optic in the CCM, and how long it takes the observatory to settle after a maneuver. Later in Sec. 5 we justify why settling time is not a driving requirement for observing efficiencies. We note, as anticipated, the tolerance ranges for M1 and M2 are tighter compared to other optics in the path, indicating their higher sensitivity in the system. Therefore, the requirement of the spacecraft attitude performance coupled to the boom actuator performance of the PRM is to reside inside the tolerance range of M1 after settling time. M1 then passes the beam through six additional CCM optics, including an asphere (M2), a low-order deformable mirror (M3), a folding mirror (M4), a second asphere (M5), a flat field scanner/3-DoF compensator (M6), and a fast steering mirror (M7), thus closing on end-to-end pointing alignment and stability at the observatory level. *This goal is met when the optical system achieves diffraction-limited performance at 30 μm (lambda) with margin thus delivering >0.8 Strehl Ratio* (see Kim et al., referenced above in this section).

*3.4.1 Structural Vibration Analysis*

The *SALTUS* induced vibration analysis leverages prior high-fidelity models of the spacecraft bus. This model was coupled with a low fidelity model of the Payload. The Payload was modeled with lumped masses to mimic M1, the truss, and boom (the M1 lumped mass), and the CCM (the CCM lumped mass) as well as springs at the connections to the instrument deck to mimic their estimated first structural modes, as shown in Table 9. In total, the Finite Element Model (FEM) used for this effort has >9 million nodes and elements. This analysis does not include deployed SM dynamics – we anticipate the low mass of each SM layer (Table 11) coupled with the low frequency first structural mode of the deployed layer will have negligible effects on pointing performance. However, the dynamics of each layer will be included in future analyses once the end-to-end observatory architecture is further refined. This assumption will therefore be revisited in Phase A. Northrop Grumman's induced vibration analysis technique has been used on multiple programs including ICESat-2, Landsat-8/-9, and JPSS-2. This method utilizes high fidelity RWA models supplied by the vendor that have been correlated to test data. Both structural modes of the model have been correlated, but also the disturbance tones including primary wheel imbalance tones and bearing disturbances. Bearing disturbances have been modeled up to 63 times the RWA wheel speed to ensure any low wheel speed excitations are correctly captured. This model also includes the appropriate rotor gyroscopic coupling as well as a uniform structural damping value for all structural modes at 0.25%. The modeling technique accurately captures the disturbances as an internal force between the bearings and the RWA housing.

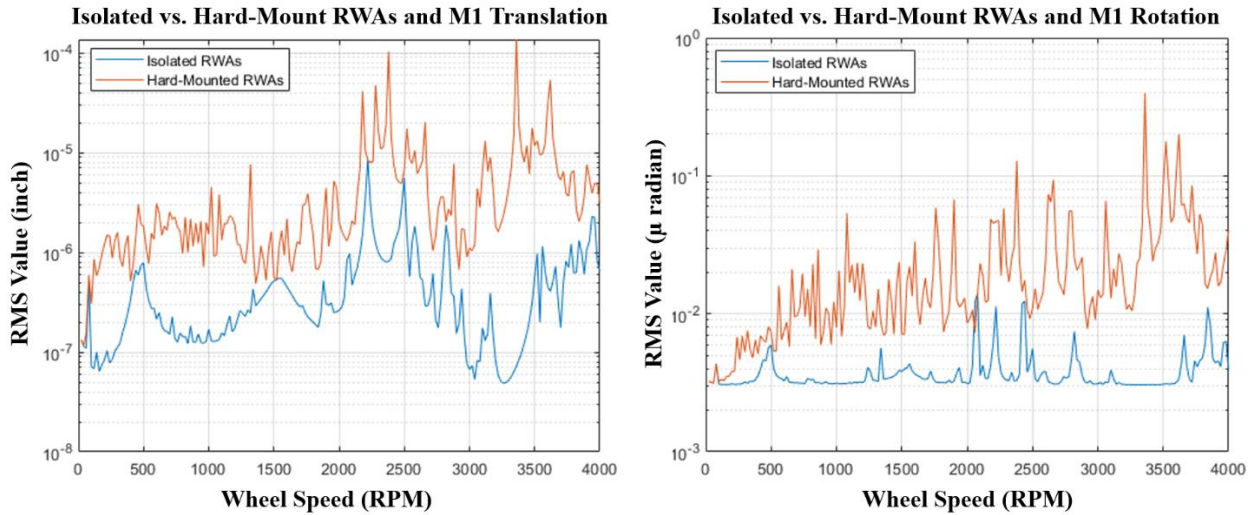

**Fig. 9 (a)** *SALTUS* observatory isolated versus hard-mounted RWAs induced vibration RMS (1σ) translational displacements of M1. **(b)** *SALTUS* observatory isolated versus hard-mounted RWAs induced vibration RMS (1σ) rotational displacements of M1.

The induced vibration analysis considered two configurations: 1) RWAs "hard-mounted" to the bus, and 2) RWAs isolated via a passive mechanical isolation system that leverages high TRL systems (e.g., Chandra and JWST). For the induced vibration analysis, the RWAs are spun up at the same wheel speed in 20 RPM increments up to 4,000 RPM, providing a conservative estimate with all the energy concentrated at the same tones. Note, the maximum allowable wheel speed (outside of safe mode) for *SALTUS* is 3,000 RPM which provides 50% margin. The frequency range of interest for this analysis was from 0.01 to 500 Hz. Both translational and rotational displacements were reported at the M1 lumped mass node. The response at the M1 lumped mass node was multiplied by a Model Uncertainty Factor (MUF) to account for model uncertainty and program maturity. The selected MUF is consistent with the concept development phase, and the amplitudes will be reduced as the design matures. The residual sum of squares for the scaled 3 DoF for translations or rotations were calculated to provide a total translational or rotational motion performance of the system. These performance curves were then used to calculate the RMS by finding the area under the curve and reporting $1\sigma$ values. We show these results in Fig. 9.

*3.4.2 PRM Settling Time After Maneuvering*

A conservative bang-bang slew was performed to estimate the initial M1 pointing error from a maneuver. The max torque from one RWA is applied about the spacecraft X axis for 50 seconds, then it is applied in the opposite direction to bring the observatory to rest. This maneuver results in an ~6° slew about the X axis. Once the body is at rest, the motion of M1 is reported resulting in an initial displacement of 0.00015 radians. From this initial displacement, various structural damping values were assumed and using the log decrement method the displacement is propagated forward in time. As seen in Fig. 10, the M1 rotational displacement as a function of time can be seen for various given structural damping values. For bare metallic or composite structures, a value of 0.15% to 0.25% is typically used, and this range in captured in Fig. 10. However, this is highly conservative and does not include the benefits of other components such as thermal blanketing and harnessing, which all contribute to additional damping. Further, we have an on-orbit point of reference for a large deployable structure from SMAP and its 6-m RBA (Table 8). This included a large boom with 2 locked out hinges, like *SALTUS*. The first two primary modes values were measured at 0.5% and 0.7%, respectively.

As noted above, we assumed a very conservative approach for the values and assumptions in settling time estimates. Additional methods can be employed to further mitigate settling time and pointing error and will be investigated later in the program. These include leveraging a slew profile that smooths the transitions to limit the system jerk, a fly-cast maneuver where the ACS system actively removes energy from the system by counteracting excitation, and other structural methods to introduce damping (such as damping tape or tuned mass dampers). We discuss settling time as it pertains to on-sky efficiency in Sec. 5.2.1.

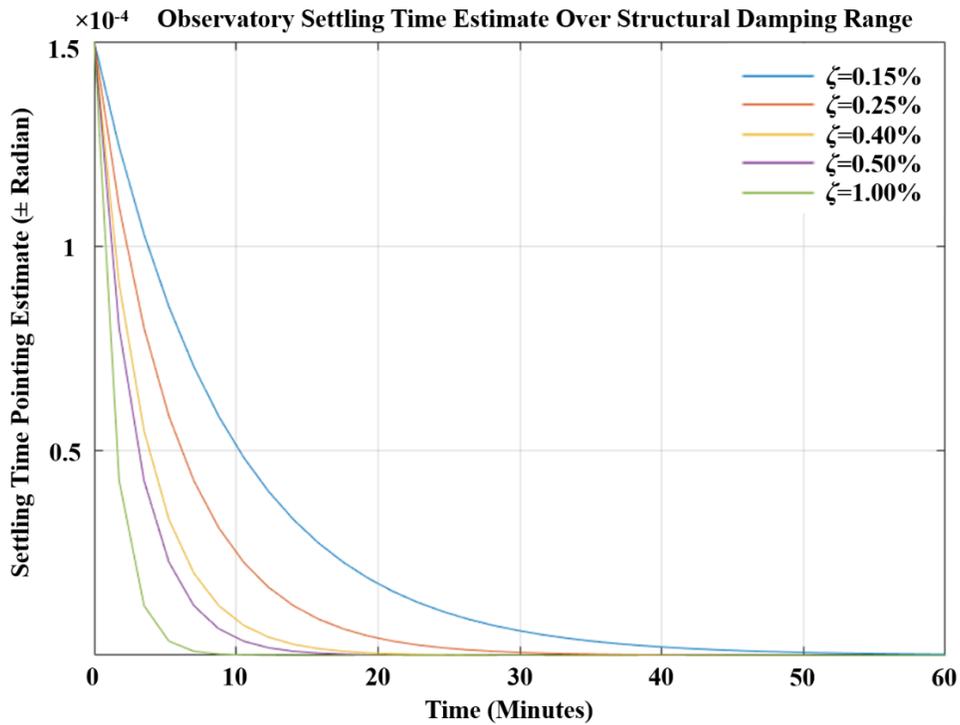

**Fig. 10** Rotational displacement of M1 as a function of time for a range of structural damping values.

In the PRM design, we also considered M1's stability with respect to M2 and find this to be primarily translation-driven by tip/tilt mechanisms in the boom and the wrist hinge – see Fig. 8 (f). This stability was assessed via FEM of the deployed PRM, assuming a 15-m truss diameter (to hold the 14-m M1), three boom sections, and four single axis actuators. Each boom section and hinge actuator was measured separately, and then combined analytically. Note, the analysis captures at minimum one axis of rotation at each of the four-hinge locations that are typically not parallel to any other axis of rotation. These locations are the root hinge between the boom and spacecraft, two intermediate hinges on the boom, and the wrist hinge between the reflector and boom. These are labelled in Fig. 8 (f), (i)-(iv). For a total boom length of ~17-m, the system is required to produce an M1 tip/tilt step size of 2.42 µrad (0.0001389°) and a linear step size of 0.025 mm. Table 8 shows previous demonstrated Northrop Grumman missions on orbit with similar performance, and the *SALTUS* Phase A design will refine this trade.

### 3.4.3 ACS Performance

The *SALTUS* ACS pointing knowledge, accuracy, and stability provides 1.19 arcsec (1σ), 1.20 arcsec (1σ), and 0.66 arcsec (1σ) at the root hinge, respectively, demonstrating significant margin to meet *spacecraft* requirements (180%, 736%, and 189%; see Table 7). Once a science target is acquired, the observatory offers <0.5 arcsec pointing stability over a 20-second period in between CCM fast-steering mirror (M7) adjustments. With this performance, the ACS enables M1 to capture the requested *SALTUS* science target and propagate to M2 in the CCM.

## 4   Sunshield Design

### 4.1  Concept Overview

The *SALTUS* baseline SM comprises two thin-film layers separated by ~2-m in the deployed configuration. This separation is driven by radiative cooling considerations to attain an M1 temperature of <45 K which is maintained throughout the mission lifetime. Each layer is deployed, tensioned, and structurally supported by a deployable mast system. The two sunshield layers, underlying structures, and associated deployment mechanisms are assembled onto a central composite cylinder, which sits outside the CIM, allowing the SM subsystem to be fully flight acceptance tested by NeXolve and Redwire Space, and delivered to Northrop Grumman for integration into the spacecraft with low complexity. Northrop Grumman designed CIM attachment points (discussed in Sec. 3.2.2) for compatibility with the SM integration. The SM structure also provides a stiff and stable root boundary condition for sunshield layers and serves as the primary structural load path for the system during launch.

**Table 10** *SALTUS* Sunshield Module heritage. The table contains heritage tensioned-membrane systems developed by NeXolve and Redwire Space. Note, *SALTUS* material thickness is driven by micrometeoroid considerations. The Solar Cruiser development is highlighted in green beside *SALTUS* (blue) as the most recent flight heritage. (1) Layer 2 area of ~1,000 m² from ~50 m × ~20 m dimension. Layer 1 area estimated at ~931 m² from ~48.5 m × ~19.2 m.

| Parameter | S4 | Nano sail-D | JWST SMA | JWST MCA | NEAScout | Solar Cruiser | *SALTUS* |
|---|---|---|---|---|---|---|---|
| **Summary System Features** | | | | | | | |
| Architecture | 4-quad | 4-quad | 5-layer | 4 covers | Monolithic | 4-quad | Monolithic |
| Years Active (*delivered) | 2005- | 2008-2011 | 2016* | 2016* | 2019 | 2019-2026 | -- |
| Quads/Layers | 4 | 4 | 5 | 1 | 1 | 4 | 2 |
| Quad / Layer Shape | Right triangle | Right triangle | 6 sided | Rectangular | Rhombus | Right triangle | Rectangular |
| Deployed Surface Shape | Flat | Flat | Doubly curved | Singly curved | Flat | Flat | Flat |
| Total Surface Area | 400 m² | 10 m² | 145 m2 – 166 m² | >20 m² each | 86 m² | 1,653 m² | ~2,000 m² (1) |
| Quad / Layer Surface Area | 100 m² | 2.5 m² | 145 m² – 166 m² | >20 m² each | 86 m² | ~413 m² | ~1,000 m² (1) |
| Film Material | CP1 | CP1 | Kapton | Kapton | CP1 | CP1 | CP1 |
| Material Thickness | 2.5 µm | 2.5 µm | 1 mil, 2 mil | 1 mil | 2.5 µm | 2.5 µm | 12.7 µm |
| **System Surface Features** | | | | | | | |
| Coating | 1000 Å VDA | 1000 Å VDA | VDA or Si | VDA or Si | 1000 Å VDA | 1000 Å VDA | 1000 Å VDA |
| Ripstop | Encap thread | Encap thread | TSB | TSB | TCP1 | TCP1 | TCP1 |
| Electrical Jumpers | -- | Yes | Yes | Yes | Yes | Yes | Yes |
| Edge Reinforce'mt | -- | Yes | Yes | Yes | Yes | Yes | Yes |
| Corner Reinforce'mt | Yes | Yes | Yes | Yes | Yes | Yes | Yes |
| Install Method | Adhesive bond | Tape | 2-part adhesive | TSB | PSA, resin bond | PSA, resin bond | PSA, resin bond |
| Seam & Rip-Stop Install. | Thermal | Thermal | TSB | TSB | Resin bond | Resin bond | Resin bond |
| **System Packaging & Deployment** | | | | | | | |
| Folded / Stowed / Packaged | Z-fold, rolled to 4 spools | Z-fold, rolled to 1 spool | Z-fold and stacked | Flat | Z-fold, rolled to 1 spool | Z-fold, nested, roll to 1 spool | Each z-folded, wrapped to 2 spools |
| Packaged Restraints / Tear Aways | None | Bumper | MRDs | MRDs | Restraint, tear aways | Restraint, tear aways | Restraint, tear aways |
| Deployment Approach | Pull out via booms | Pull out via booms | Complex | Rel. mech., rolled | Pull out via booms | Pull out via booms | Pull out via booms |
| Tensioned Method | 3-corner pull via ATK boom | 3-corner pull via metallic TRAC boom | 6-point corner pull | -- | 4-corner pull via metallic TRAC boom | 3-corner pull via composite TRAC boom | 4-corner pull via composite TRAC boom |

The SM design heavily leverages several solar sail designs spearheaded by NeXolve and Redwire Space. Most recently, the Solar Cruiser[6,7] solar sail development program has matured a 40 m$^2$ aperture to TRL-8 through full flight qualification testing. The deployable masts for Solar Cruiser are 30-m long, comparable in length to the mast design for *SALTUS*. These masts, discussed in more detail in Sec. 4.3.1, have been designed and analyzed to accommodate the structural loading anticipated for *SALTUS*. A full mast test campaign has resulted in structural performance verification and correlated models [11]. In this section, we discuss heritage and TRL, the SM structure, as well as stowed and deployed configurations, and the deployment sequence. Finally, we expand on key design considerations unique to *SALTUS* that drove design decisions such as the sunshield area, shape, and positioning with respect to M1.

*4.2 Heritage and TRL*

The two-layer *SALTUS* sunshield draws strong heritage from NeXolve's experience in designing and building gossamer membrane systems such as the JWST sunshield, the solar sails on missions like NEAScout and Solar Cruiser, and deorbiting and dragsail devices as deployed on Nanosail-D. We have included six exemplar developments and deployments in Table 10 which includes comprehensive detail of system features, surface (material) features, as well as packaging and deployment designs. In the right-most column of Table 10, we show the current *SALTUS* baseline as it relates to other programs, most notably the Solar Cruiser mission highlighted in green, currently in successful operation with significant crossover in design with *SALTUS*. Indeed, the selected materials are based primarily on thermal performance coupled with mechanical considerations, and the design would likely use Kapton (JWST) or CP1 (Solar Cruiser) for the film material, with optical coatings comprised of vapor-deposited aluminum (VDA, typically 1000 Å) and high-emissivity coatings (Si) on the back side for increased heat rejection. These material and coating choices will be down-selected pending a full system end-to-end thermal analysis in Phase A. The *SALTUS* design therefore takes advantage of tried and tested materials and processes developed for successful programs, resulting in a low risk approach while meeting thermal and mechanical requirements. Key parameters included in Table 10 are discussed in Sec. 4.3.1.

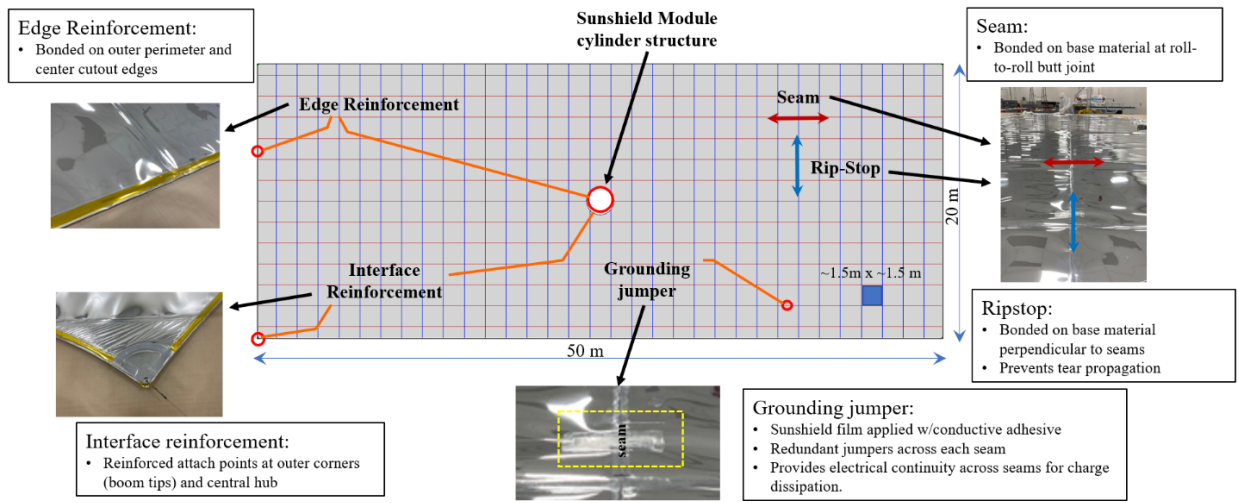

**Fig. 11** *SALTUS* sunshield layer features and layout. Image credit: NeXolve. Northrop Grumman obtained permission from NeXolve to include this graphic.

## 4.3 Sunshield Module Structure and Deployment Sequence

### 4.3.1 Structure

Figure 11 shows the typical features and layout that will be employed for SM layers. Roll widths vary depending on the material selected (ranging 36" - 60") and will drive the number of seams required to span the 20-meter sunshield width. Seams will be parallel to the long edge with Z-folding perpendicular to these seams as the sunshield layer membranes are manufactured. Seaming processes are dependent on the material selected; some options include solvent bonding (CP1), thermal spot bonding (Kapton), or the use of silicone adhesives (e.g., Nusil). Seams typically have grounding "jumpers" installed to ensure that the sunshield membrane is electrically continuous and can dissipate any charge that may build up through a ground connection with the spacecraft.

Perpendicular to the seams are ripstops, typically spaced to create a square grid pattern with the seams. Ripstops are to limit propagation of a tear should one occur. The installation of ripstops is similar to seams. Trimmed edges, found at the outer perimeter and at the circular cutout where the SM's cylindrical structure resides (red circle in the center of Fig. 11, and labeled as "Central Structure" in the stowed configuration in Fig. 12) are reinforced to prevent tears and to distribute tensioning load across the sunshield membrane. Constructions vary from using Kapton tape (with a cover overlay to mitigate adhesive leach out) to bonding on strips of the base material to "double up" the thickness at this edge. Corners have additional reinforcements where the grommet attachment point is located, typically a stack-up of stainless steel (3-5 mil depending on membrane tensioning loads) and additional covers for optical purposes, and grounding features. Connection to the mast can be through a small clevis/pin attached to the grommet for ease of installation/removal to the cable/tensioning system at each of the boom tips. Grounding of the membrane to the spacecraft can be done here or at the CIM interface if attached (or both).

Triangular Rollable and Collapsible (TRAC) masts are constructed from two thin gage High Strain Composite (HSC) tape springs that are bonded along one edge to form a triangular cross-section. This geometry can be flattened and then rolled onto a hub for stowage (see Fig. 12). 30-m TRAC masts, the same length required for the *SALTUS* sunshield, have been demonstrated and matured to TRL-6 as part of the Solar Cruiser solar sail system. The TRAC mast provides very high bending stiffness for a given flattened height. In other words, the mast provides excellent performance for a minimized system mass and stowage volume. Other rollable cross-sectional geometries, such as the Collapsible Tubular Mast (CTM), could also be considered for the masts. While the CTM cross-section provides substantially higher torsional stiffness and stability than the TRAC variant, further development work would be needed to fabricate and demonstrate the CTM at the required length and stiffness needed for *SALTUS*. However, the CTM technology is being matured as part of NASA Langley's ACS3 mission and could certainly be a viable option for *SALTUS*, given the timeframe for the mission. Apart from cross-sectional geometry, these two mast technologies are very similar with respect to the composite materials used, as well as their spooling and deployment behavior, and are therefore considered interchangeable within the sunshield subsystem. The composite TRAC masts are manufactured from carbon fiber reinforced polymer (CFRP) composite materials. The high-modulus carbon fibers are predominantly oriented in the longitudinal axis, providing optimized compression, and bending stiffness at a minimized mass. Furthermore, the composite laminate architecture is optimized for near-zero CTE to ensure on-orbit shape stability. The *SALTUS* SM team has extensive experience with composite mast design and further optimizations can be considered if necessary.

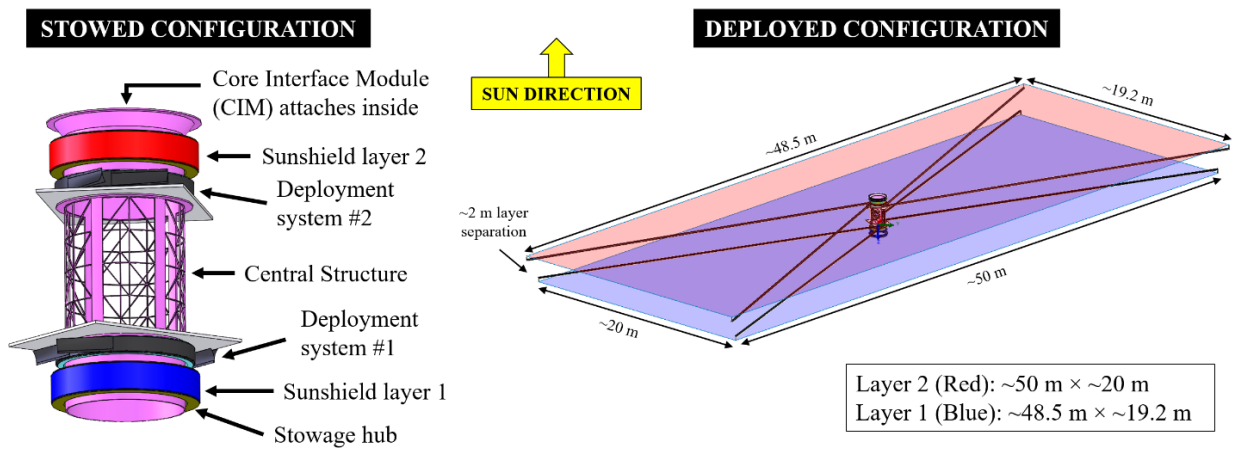

**Fig. 12** *SALTUS* sunshield stowed and deployed configurations. The stowed configuration shows a zoomed in image of the Sunshield Module (SM) which sits outside, and attaches to the CIM. The smaller ~48.5 m × ~19.5 m layer 1 deploys first, followed by the larger ~50 m × ~20 m layer 2, which avoids a cryogenic deployment of the second layer to deploy. Layer 2 will be the last major deployment of the *SALTUS* observatory.

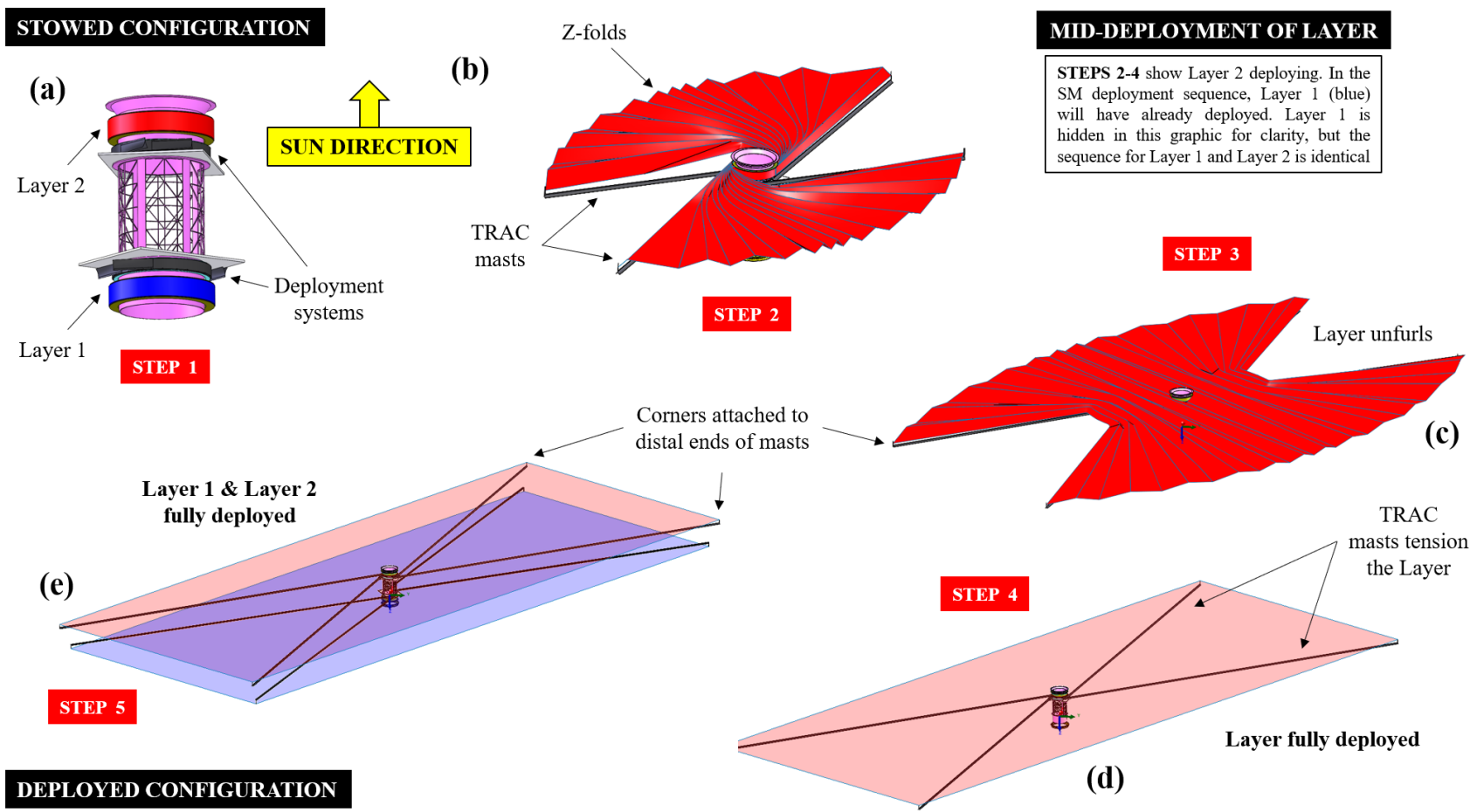

**STOWED CONFIGURATION**

**(a)**

SUN DIRECTION

Layer 2

Deployment systems

Layer 1

**STEP 1**

**(b)**

Z-folds

TRAC masts

**STEP 2**

**MID-DEPLOYMENT OF LAYER**

**STEPS 2-4** show Layer 2 deploying. In the SM deployment sequence, Layer 1 (blue) will have already deployed. Layer 1 is hidden in this graphic for clarity, but the sequence for Layer 1 and Layer 2 is identical

**STEP 3**

Layer unfurls

**(c)**

Corners attached to distal ends of masts

**Layer 1 & Layer 2 fully deployed**

**(e)**

**STEP 5**

**DEPLOYED CONFIGURATION**

TRAC masts tension the Layer

**STEP 4**

**Layer fully deployed**

**(d)**

Fig. 13 *SALTUS* sunshield deployment sequence. **(a)** Stowed configuration showing each sunshield layer and deployment system location. The stowed layers, deployment system, and TRAC masts are co-spooled on a stowage hub. **(b)** TRAC masts are deployed using a single brushless DC motor mechanism. **(c)** The deployment mechanism continues to unfurl the layer. **(d)** Tension is driven into the layer on full deployment. The corners of each layer are mechanically attached to the distal ends of each mast. **(e)** STEPS 1-4 are repeated for the second layer ensuring a non-cryogenic deployment for both layers. Once both layers are deployed, all major observatory deployments are complete and the cryogenic thermal environment begins to stabilize to <45 K.

*4.3.2 Deployment Sequence*

The sunshield is the last major deployment for the observatory, which follows solar array and HGA deployments and PRM deployment. Sunshield layer 1 (the smaller layer closer to the PRM) is deployed first, thus ensuring that no preceding subsystem/component deployment is a cryogenic deployment, which would require additional cryogenic development and testing. Layer 2 is the final deployment in the sequence. We describe the sequence in reference to Fig. 13, as follows:

- **<u>STEPS 1-3</u>**: STEP 1 shows the SM in a fully stowed state. The rectangular sunshield layers are pleated (i.e., z-folded) in the long axis, and then spooled onto a cylindrical hub for stowage. Adjacent to each stowed layer is its associated deployment system, comprised of four TRAC masts that are co-spooled on a stowage hub, and are used to deploy and tension each layer. STEPS 2-3 show the mid-deployment of layer 2 following TRAC masts deployment from their stowage hubs. As noted in the figure, layer 1 will deploy first and has been hidden for clarity, but the sequence is identical. Layers can be deployed separately via independent deployment systems (a single brushless DC (BLDC) motor), which occurs in <60 minutes and utilizes <20 W of power. These single motorized mechanisms contain a variety of features to stabilize the system for launch, control boom deployment in-flight, as well as to provide the necessary root boundary constraints for deployed structural performance. The corners of the stowed rectangular layer are mechanically attached to the distal ends of the masts. Figure 14 (c) illustrates a four-mast TRL-6 deployment.

- **<u>STEPS 4-5</u>**: Once the masts are deployed, the bearing-mounted sunshield layer is fully unfurled. Tension is driven into each layer upon full deployment, resulting in a planar layer that meets key performance parameters for flatness and deployed first mode natural frequency. Once STEP 4 is completed for layer 1, STEPS 1-4 are repeated for layer 2 resulting in the full 2-layer sunshield deployment, as shown in STEP 5.

The deployment mechanism we selected for *SALTUS* will heavily leverage the Sail Deployment Mechanism (SDM) designed for the Solar Cruiser solar sail[6,7]. The Solar Cruiser SDM was designed to deploy four 30-m long TRAC booms in a square configuration and has been matured to TRL-6 as part of the Solar Cruiser solar sail flight qualification campaign.

**Table 11** *SALTUS* Sunshield Module component masses. The total MEV mass is captured under "Payload" (Table 3).

| Component | Qty | CBE Unit Mass (kg) | Contgy. (%) | MEV (kg) |
|---|---|---|---|---|
| Sunshield Layer 1 | 1 | 18.3 | 25% | 22.9 |
| Sunshield Layer 2 | 1 | 19.6 | 25% | 24.5 |
| Hub Hardware | 2 | 1.5 | 25% | 3.8 |
| Booms | 8 | 2.8 | 25% | 28.0 |
| Boom Deployer Hardware | 2 | 20.0 | 25% | 50.0 |
| Central Cylinder | 1 | 8.0 | 25% | 10.0 |
| Misc | 1 | 5.0 | 25% | 6.3 |
| **TOTAL** | | | **25%** | **145.4** |

*4.4 Key SALTUS Design Considerations*

The initial sizing of the sunshield was driven by the maximum achievable FOR of the *SALTUS* observatory using current state-of-the art capabilities in the design and manufacturing of tensioned-membrane architectures. As previously noted, the *SALTUS* sunshield heavily leverages the recent Solar Cruiser solar sail development program which has brought a 40 m × 40 m deployable membrane to TRL-6, involving a full environmental test campaign. A full-scale quadrant of the Solar Cruiser solar sail is shown in Fig. 14 (a).

Due to *SALTUS* CONOPs, the sunshield transverse dimension can be much smaller than the longitudinal dimension, which is advantageous to system mass. A 20-m transverse dimension was therefore selected, which is the maximum membrane dimension that can be manufactured using existing GSE and facilities at NeXolve. The resulting rectangular architecture represents a slight departure from Solar Cruiser's square architecture; however, the masts and membranes used in Solar Cruiser are similar in size to what is envisioned for *SALTUS* and therefore do not present scaling challenges. The most substantial modification will involve adjusting mast angles within the deployment mechanism to achieve the rectangular geometry. Importantly, the *SALTUS* design does not involve lengthening the masts beyond 30-m (demonstrated for Solar Cruiser), or widening the membrane beyond 20-m, which means that existing GSE and processes can be utilized, as shown in Fig. 14 (b).

As the *SALTUS* mission concept is refined in Phase A, detailed modeling, and analysis of the sunshield system, both structural and thermal, may lead to additional design trades. As an example, the performance of the telescope may greatly benefit from the addition of one or more membrane layers, aiding in the protection against MMs. Additionally, alternative optical coatings having improved emissivity and improved thermal rejection capabilities may be investigated.

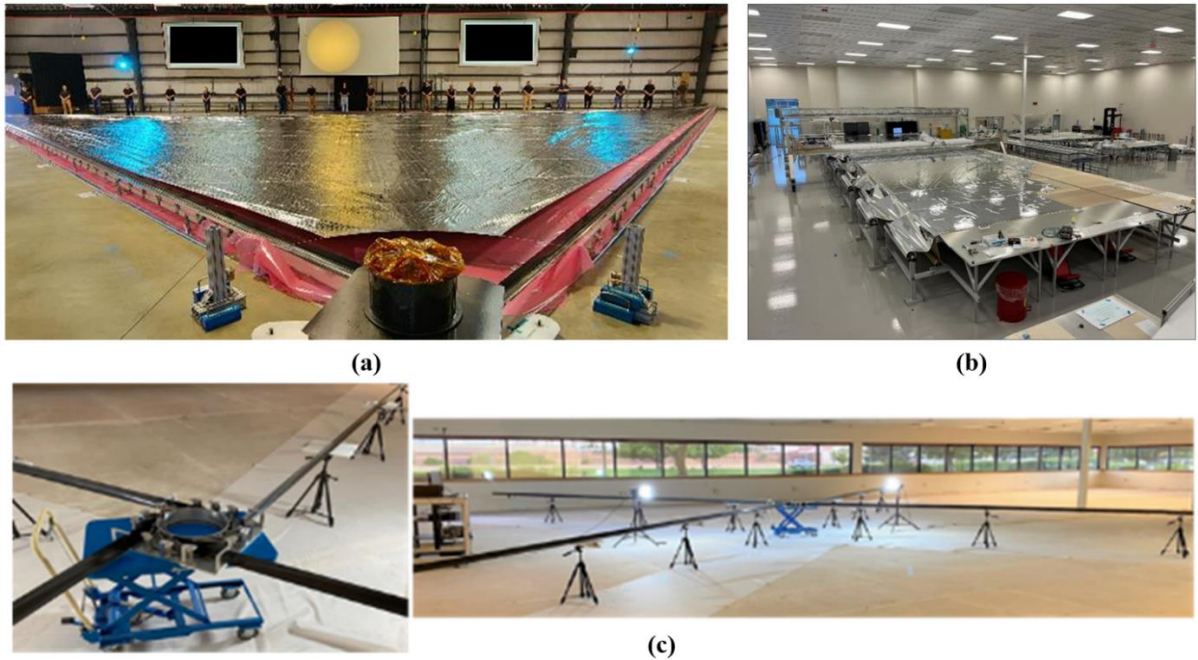

**Fig. 14 (a)** Single deployed quadrant of the Solar Cruiser sail system utilizing 30 m TRAC masts with a side dimension of 40-m. **(b)** Membrane manufacturing facility at NeXolve which is capable of producing sunshield layers to meet the *SALTUS* area. **(c)** A four-mast TRAC deployment system matured to TRL-6. Image credit: NeXolve (a, b), Redwire Space (a, c). Northrop Grumman obtained permission from NeXolve and Redwire Space to include this graphic.

## 5 Lifetime: A Two-Consumable Architecture

### 5.1 The Propellant System

#### 5.1.1 Overview

The *SALTUS* propellant tank has been sized to carry a total hydrazine propellant mass of 1093.7 kg MEV. ~40% of this budget is required to reach the *SALTUS* orbit via impulsive burn and X/Y attitude control maneuvers, corresponding to the $\Delta V$ budget in Table 2. The remaining ~60% is required for RWA MUs (~40%) and station keeping (~20%). Our calculations show that MU is dominated by SRP and inertia, both unique to *SALTUS*, and thus we consider MU maneuvers to be the dominant propulsion mode that limits mission lifetime when considering the "propellant consumable" in the two-consumable architecture (the other consumable being ICS, discussed in Sec. 5.2). SRP accumulates across the *SALTUS* FOR over the mission lifetime (including CG migration) which scales directly from the total sunshield larger layer area of ~1,000 m². Additionally, as shown in the table in Fig. 5 (b), the observatory produces very large inertial moments (>80,000 km·m² in a deployed configuration at BOL) thus requiring large RWAs to meet acceptable slewing performance across the FOR. These factors require significantly more propellant mass for MUs than other conventional space missions and ultimately define lifetime. We consider MU calculations to be a bounding case because the analysis does not capture strategies that minimize momentum accumulation, e.g., strategic observation planning for efficient target selection, which could extend mission lifetime even further. Currently, the propellant subsystem provides positive margin at the 5-year mission baseline with a tank capacity of 50.1% (Table 7). In this subsection, we expand on *SALTUS* on-sky efficiency, MU, and discuss migration of the static margin (i.e., distance between CG and Center of Pressure, CP) from BOL to EOL.

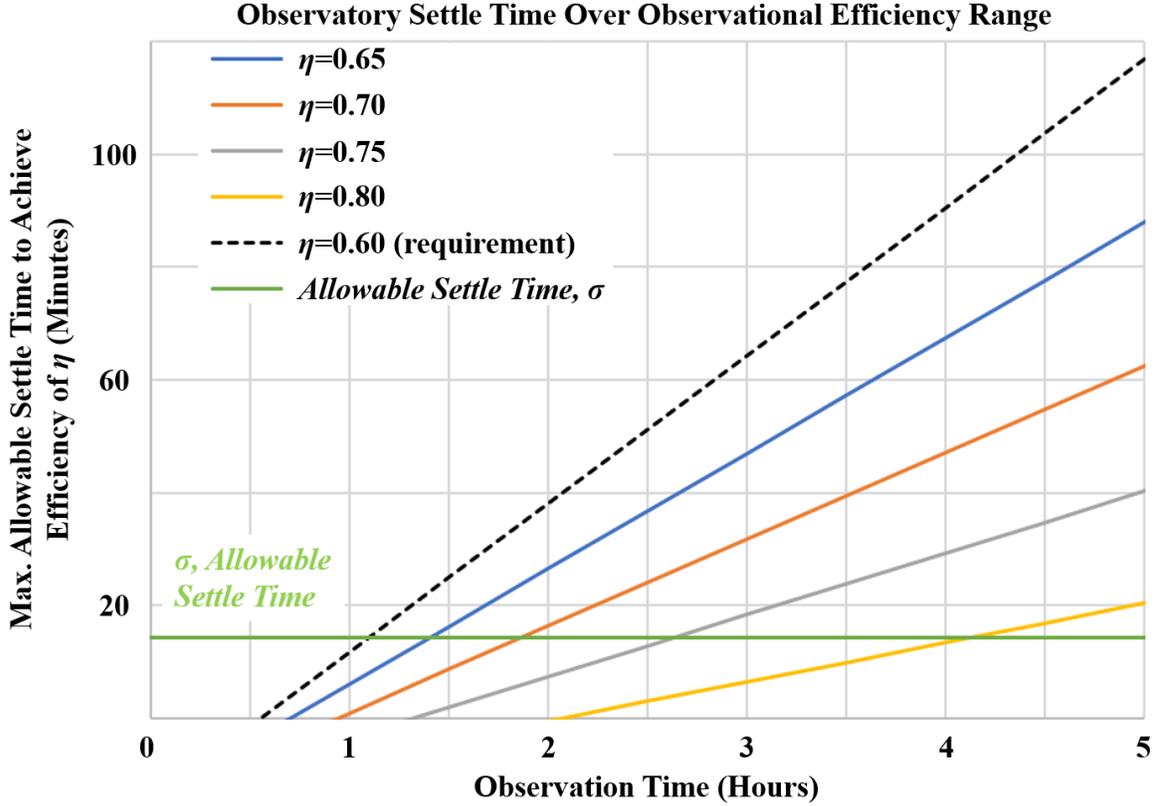

**Fig. 15** *SALTUS* maximum allowable settle time to achieve a given observing effiency as a function of observation time. For shorter observations, the mean efficiency is greater than science mission requirements. Given the requirement of $\eta$=0.60, the mean observation time can be as short as 1h (current mean observation time is 5h CBE) and *SALTUS* will still meet this requirement.

*5.1.2 Operational On-Sky Efficiency*

The operational efficiency, $\eta$, of *SALTUS* for the GTO program is the ratio of time observing to the total time (also characterized as 1 minus the fraction of time spent not observing), as follows:

$$\eta = \frac{time\ observing}{all\ time} = \frac{(all\ time) - (time\ not\ observing)}{all\ time} \tag{1}$$

This $\eta$ requirement for *SALTUS* is 60%. Although the GO observing program is currently unknown, the GTO's performance discussed herein is directly applicable to the GO program. *SALTUS* will not observe during slew and/or settling time, MUs, ground contacts, or observatory safe modes. We carried out a parametric assessment assuming a slew, MU, and settle before and after an observation, which shows the determination of observational efficiency, $\eta$, can be reduced to a single parameter – allowable settle time, $\sigma$. Note, per Table 5, the worst-case power mode does in fact assume an observation during ground contacts for conservatism in power sizing, and although observatory does have the ability to accommodate this mode, it should not affect the outcome of $\eta$ or $\sigma$. The (*time not observing*) variable in Eq. (1) can be broken up into the following components: the observatory overhead, $\gamma$; the percentage of time lost to communications, $n(S + \sigma + L_T)/7.24$; the percentage of time lost to MU, $((24/\delta) \cdot t_d)/24$; and the percentage of time lost to spacecraft safe mode from flight actuals, $W_{sc} = M/365$, where $M$ is the total time the spacecraft spends in safe hold mode over 1 year. Equation (1) now becomes:

$$\eta = \frac{24 - \left(\frac{24}{L} \cdot (S + \sigma)\right)}{24} - \gamma\left(\left(\frac{n(S + \sigma + L_T)}{7.24}\right) + \left(\frac{\frac{24}{\delta} \cdot t_d}{24}\right) + W_{sc}\right), \qquad (2)$$

where $L$ is the length of an observation, $S$ is the slew time, $\sigma$ is the allowable settle time and is a *variable* to be solved for in this derivation, $n$ is the number of communications contacts per week, $L_T$ is the total link time, $\delta$ is the MU interval, and $t_d$ is the time taken to unload the RWAs. Therefore,

$$\eta = 1 - \left(\frac{S + \sigma}{L}\right) - \gamma\left[\left(\frac{n(S + \sigma + L_T)}{7.24}\right) + \left(\frac{t_d}{\delta}\right) + W_{sc}\right], \qquad (3)$$

$$\eta = 1 - \left(\frac{S + \sigma}{L}\right) - \left(\frac{\gamma \cdot n(S + \sigma)}{7.24}\right) - \left(\frac{\gamma \cdot n \cdot L_T}{24}\right) - \left(\frac{\gamma \cdot t_d}{\delta}\right) + \gamma \cdot W_{sc}, \qquad (4)$$

$$-\left(\frac{\sigma}{L}\right) + \left(\frac{\gamma \cdot n \cdot \sigma}{7.24}\right) = 1 - \left(\frac{S}{L}\right) - \left(\frac{\gamma \cdot n \cdot S}{7.24}\right) - \left(\frac{\gamma \cdot n \cdot L_T}{24}\right) - \left(\frac{\gamma \cdot t_d}{\delta}\right) - \gamma \cdot W_{sc} - \eta \qquad (5)$$

Solving for the allowable settle time, $\sigma$, we find:

$$\sigma = \frac{1 - \left(\frac{S}{L}\right) + \left(\frac{\gamma \cdot n \cdot S}{7.24}\right) - \left(\frac{\gamma \cdot n \cdot L_T}{24}\right) - \left(\frac{\gamma \cdot t_d}{\delta}\right) - \gamma \cdot W_{sc} - \eta}{\left(\frac{1}{L}\right) + \left(\frac{\gamma \cdot n}{7.24}\right)} \qquad (6)$$

Evaluating Eq. (6) for a range of $L$ and potential values of $\eta$, we use the following analysis to determine input parameters. We assume *SALTUS* targets are uniformly distributed over the FOR. The observatory is capable of slewing 180° in ~30 minutes via six RWAs (see Sec. 3.2.7) at 50% max allowable wheel speeds of 3,000 RPM (providing 50% margin), and by using the maximum inertia in the deployed configuration from Fig. 5 (b), we find a mean slew $S$~15 minutes. $L_T$ is a total of 7h per week distributed per the science CONOP, contributing 4.2% of a day. $\delta$ is expected to occur at the shortest interval, i.e., the worst-case attitude and at BOL when CG-CP is a maximum, which is every ~4 hours and take ~5 minutes to complete. This contributes 2.1% of a day. We combine this with an historical average for LEOStar-3 safe modes duration, $W_{SC}$, at 3 days per year, or 0.8% per day. If the mean observation, $L$, is 4 hours (currently the mean is estimated to be ~5 hours, but we use 4 hours for added conservatism), there are 6 observations per day thereby resulting in 5.2% of a day in slewing modes. These activities account for 12.3% of the time. We added $\gamma$=25% observatory overhead (i.e., 12.3% × $\gamma$, where $\gamma$=1.25) to capture uncertainties thereby providing extra contingency to the design process, yielding 15.4% of time lost per day. This results in $\eta$=84.6% observational efficiency, well in excess of the $\eta$=60% requirement, as shown in Table 12.

Solving Eq. (6) with the above inputs yields an allowable settle time, $\sigma$, of 13 minutes CBE (~1% of a day), as indicated by the green horizontal line in Fig. 15 which shows the range of compliance against the $\eta$=60% efficiency requirement. We note to meet $\eta$=60%, <27.6% of the day must be spent settling, or 4.6% of a day per observation (~1.1 hours). Effectively, this means with $\sigma$=13 minutes, $L$ can be as short as 1h (recall, the current mean $L$=5h CBE), indicating that the *SALTUS observing efficiency is highly insensitive to the settle time*.

**Table 12** *SALTUS* on-sky efficiency key & driving variables. (1) 7h per week distributed per CONOP. (2) Expected to occur at shortest interval, worst case BOL attitude. (3) From LEOStar-3 flight actuals. (4) CBE observation time is 5 hours; 4 hours used for added buffer. Therefore, assume 6 observations per day, each requiring a slew to target. (5) Observatory overhead to capture uncertainties in the design process, which we add as an additional $\gamma$=25% to 12.3%.

| Variable | % Lost Per Day | Value |
|---|---|---|
| Communications Contacts (1) | 4.2% | 7 hours /week |
| Momentum Unload Interval (2) | 2.1% | 5 minutes /4 hours |
| Spacecraft Safe Hold Duration (3) | 0.8% | 3 days /year |
| Slewing Modes (4) | 5.2% | 6 slews /day |
| **TOTAL: (1)+(2)+(3)+(4)** | **12.3%** | **--** |
| Observatory Overhead (5) | 3.1% | 25% of TOTAL |
| **ON-SKY EFFICIENCY** | **15.4%** | **84.6%** |

*5.1.3 Momentum Unloading, Torque, and Static Margin*

The interval between MUs is dependent on separation of the CG to the CP, the observatory attitude, and the sunshield properties. The table in Fig. 5 (b) shows observatory CG locations, in meters, in X, Y, and Z with respect to the spacecraft origin, which we define to be the geometric center of the separation ring plane located at the ~aft end of the spacecraft (labeled "Sep plane" in Fig. 5 (c)). We designed CG to approximately reside at the X- and Y-centers of the GEOStar-3 primary structure. CP is located at the geometric center of sunshield layer 2 (on the spacecraft side; the 50 m × 20 m layer) at XYZ coordinates 0 m, 0 m, +3.88 m. By assessing the mass, and the XYZ location of all major *SALTUS* spacecraft components, flight system components, and Payload components, we calculated CG at launch and over the full mission lifetime (BOL-EOL). We assessed the initial CG LV dispersions and linear motion as the consumables are expended over mission life. Since we know the momentum storage capacity of the selected RWAs, we can calculate a map of MU intervals as a function of observatory pitch and roll angles, corresponding to different attitudes in the FOR.

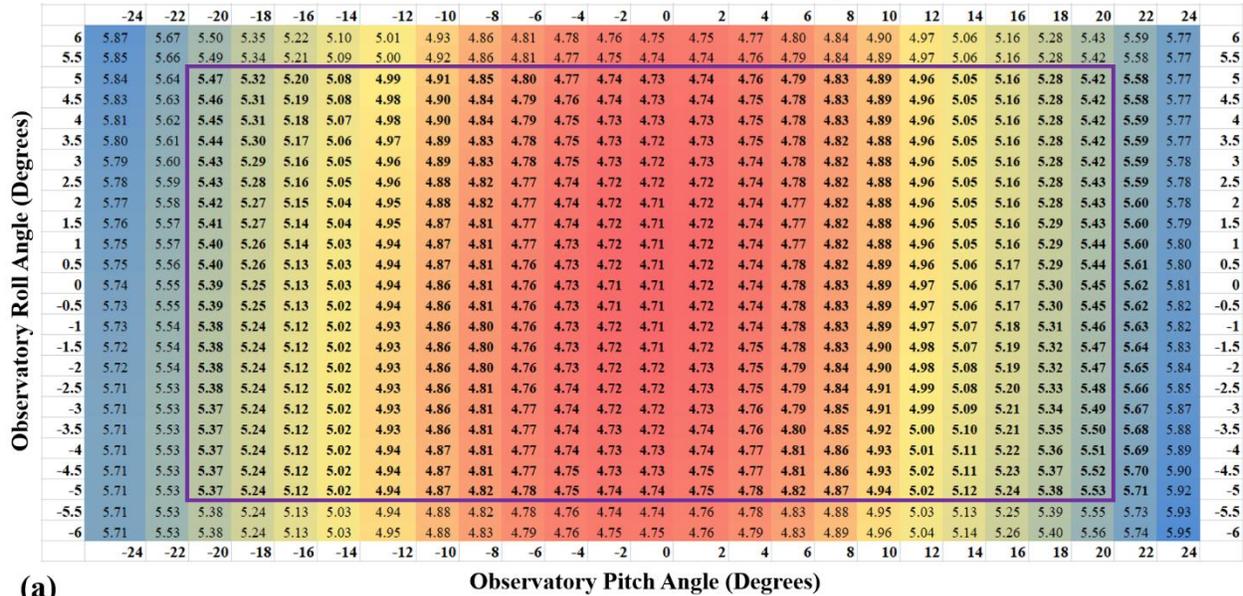

**(a)**

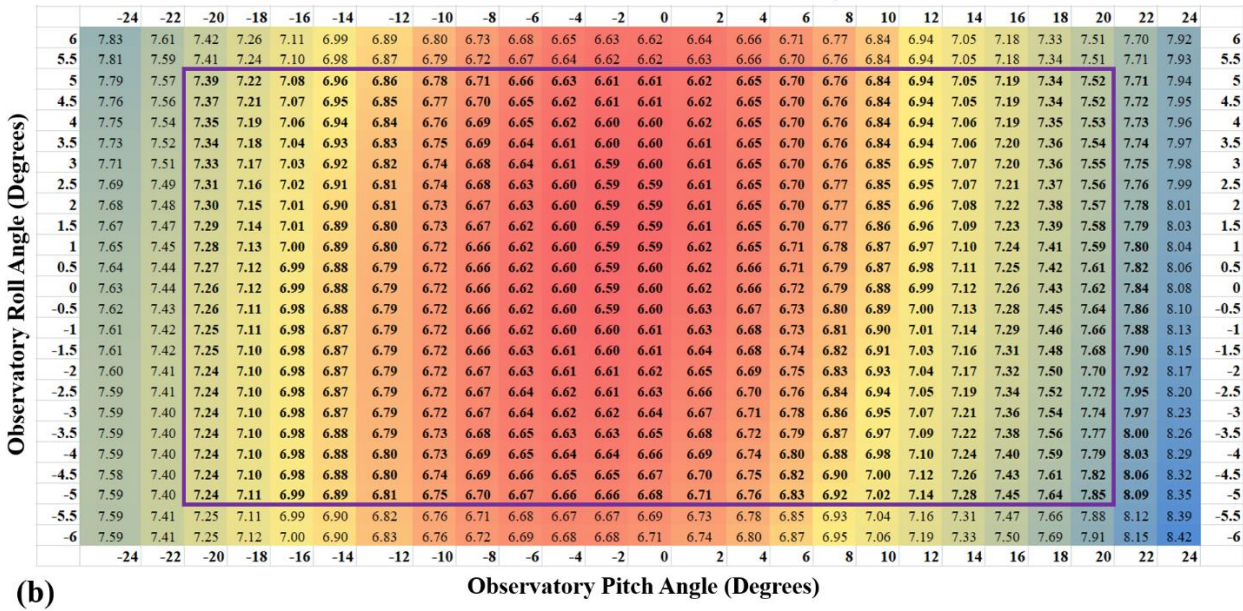

**(b)**

**Fig. 16** *SALTUS* momentum unload interval in hours as a function of observatory pitch and roll at **(a)** BOL and **(b)** EOL over the *SALTUS* FOR. Each number in a colored cell coresponds to the unload interval in units of hours. The smaller the number, the shorter the unload interval, which indicates greater momentum accumulation in the RWAs at that attitude. Intervals increase with mission lifetime as propellant is used and CG migrates closer to CP. The purple boxes in (a) and (b) highlight the *SALTUS* FOR defined by ±20º of pitch, and ±5º of roll, and show a BOL minimum unload interval of 4.71h, and an EOL minimum unloaod interval of 6.59h.

*5.1.3 Momentum Unloading, Torque, and Static Margin*

The interval between MUs is dependent on separation of the CG to the CP, the observatory attitude, and the sunshield properties. The table in Fig. 5 (b) shows observatory CG locations, in meters, in X, Y, and Z with respect to the spacecraft origin, which we define to be the geometric center of the separation ring plane located at the ~aft end of the spacecraft (labeled "Sep plane" in Fig. 5 (c)). We designed CG to approximately reside at the X- and Y-centers of the GEOStar-3 primary structure. CP is located at the geometric center of sunshield layer 2 (on the spacecraft side; the 50 m × 20 m layer) at XYZ coordinates 0 m, 0 m, +3.88 m. By assessing the mass, and the XYZ location of all major *SALTUS* spacecraft components, flight system components, and Payload components, we calculated CG at launch and over the full mission lifetime (BOL-EOL). We assessed the initial CG LV dispersions and linear motion as the consumables are expended over mission life. Since we know the momentum storage capacity of the selected RWAs, we can calculate a map of MU intervals as a function of observatory pitch and roll angles, corresponding to different attitudes in the FOR.

We show this result in Fig. 16 (a) where we have highlight in a purple rectangular box MU interval in hours at BOL as a function of the *SALTUS* FOR, as defined by ±20º of pitch and ±5º of roll. By finding the minimum MU interval in the FOR, we can identify the attitude that generates maximum momentum accumulation. As shown, the worst-case MU interval is 4.71 hours at BOL, which is less frequent than the observation length of 4 hours used in our efficiency calculation in Sec. 5.1.2. Moreover, this worst-case MU interval increases over mission life as the propellant and inflatant gas are consumed, and therefore this 4-hour MU interval is used to size the propellant mass over the *SALTUS* mission lifetime. At EOL, the MU interval minimum is 6.59 hours, as shown in Fig. 16 (b). To assess the adequacy of the propellant capability of the proposed tank arrangement, we simulated many likely *SALTUS* missions via Monte Carlo simulation, varying the fuel needed for achieving orbit under launch dispersions. We ran this Monte Carlo for a ~few hundred cases of nominal properties and compared to the 95th percentile storage capability to demonstrate adequate propellant capability of the proposed *SALTUS* design. Furthermore, we note for the MU interval calculations, we used a brute force approach for each case, where we assume the RWAs are loaded to capacity in one direction and are required to unload fully before loading in the other direction. We have therefore not considered a low torque accumulation plan which would reduce propellant even further. This calculation is thus very conservative.

The conservative approach to estimating the fuel needed for the *SALTUS* mission gives us confidence that further design analysis, even if it reveals additional propellant needs, will validate that there is enough propellant storage in our design. It is likely that with time for more simulation and analysis we can optimize our propellant storage design further.

## 5.2 The Inflation Control System

### 5.2.1 M1 Inflation

The ICS is a closed loop control system that regulates the flow of helium inflatant from four high-pressure (>4,000 psi) storage tanks on the spacecraft primary structure through gas lines on the boom to M1 and maintains the pressure within M1 to meet performance requirements over all attitudes and through the mission during science operations. The ICS maintains M1 pressure during science observations 5.1 Pa ± 5.1 mPa. The ICS consists of a large number of high TRL components from previous Northrop Grumman missions requiring pressure-regulated systems, where all ICS components have been identified by a commercial catalog number. The pressure in M1 is sensed through six strain gauges made from piezo-electric film bonded to M1 and connected in a 4-wire arrangement, located outside the optical clear aperture – see the *SALTUS* block diagram in Fig 4. The strain gauge produces a voltage proportional to film strain, which is proportional to pressure and forms the basis of the error signal for the control loop. With large gauge functions, it provides sufficient resolution and accuracy for control of M1's pressure. The baseline approach is to operate this loop as a bang-bang controller, but other control algorithms will be studied in Phase A. Also in Phase A, the sensor will be exposed to radiation and low temperature to demonstrate that the behavior of the sensor over these environments is suitable for the *SALTUS* mission.

The control function is housed in the PIE and communicates directly to the "ICS Controller" in the block diagram (Fig. 4). Based on the error signal, helium gas will either flow into M1 or be released from it. The ICS has sufficient resolution that the error between the continuous loss rate of M1 and the discrete inflow rates have a mean error of a few nmol/sec, allowing the ICS control loop to maintain pressure to within requirements. The *SALTUS* inflatant gas lasts for the required 5 years of mission life under very conservative assumptions (with an allocation of 200 kg, see Table 3) and is likely to last significantly longer under nominal conditions. Factors contributing to inflatant consumption are penetration by MM, creep induced hole growth, valve, and fitting leakage, cold to hot losses (see Arenberg et al., "Design, Implementation, and Performance of the Primary Reflector for SALTUS," *J. Astron. Telesc. Instrum. Syst.* (this issue)), and permeation[8]. The brass board ICS will be part of Phase B testing to show the combined M1 ICS functions together as a system.

### 5.2.2 Micrometeoroids (MM)

MM bombardment causes a linearly increasing leak rate with time, resulting in a time-squared dependence on the inflatant mass[8,9]. To make an accurate estimate of the area of penetrations over time, Northrop Grumman carried out a series of hypervelocity impact tests on representative membranes and produced a model of fragmentation, to accurately determine the total area damaged[8]. Reference 8 outlines these test and Arenberg et al., "Design, Implementation, and Performance of the Primary Reflector for SALTUS," *J. Astron. Telesc. Instrum. Syst.* (this issue) modify this approach to give a calculation of the total area damaged from the full MM flux for membranes of thickness and properties appropriate for *SALTUS*. As an additional contingency factor to create a worst-case expectation of MM penetration, Northrop Grumman also uses an internally developed tool, MADRAT, for the more specific calculation of system penetration including factors such as orientation, system shielding, orbital velocity, and natural anisotropies in the environmental flux. MADRAT is validated against NASA's BUMPER code[10] and has specifically included a factor for hole growth due to visco-elastic creep as part of worst-case estimates of gas consumption. Polymeric films as will be used on M1 are expected to have negligible creep as the operating temperature is low and far from the glass transition temperature[11]. The losses due to cold to hot are assumed to be incurred every other observation and taken for the full venting of 0.04 to 0.07 mols, depending on temperature. The vendor-specified valve losses for both internal and external leaks are assumed to be entirely external, then doubled as additional contingency. Assessment of the inflatant lifetime is performed with worst-case and nominal CBE parameters. The 5-year inflatant lifetime is met under worst case and nominal conditions. We refer to the reader to Arenberg et al., "Design, Implementation, and Performance of the Primary Reflector for SALTUS," *J. Astron. Telesc. Instrum. Syst.* (this issue) for an in-depth discussion of M1 design, properties, and performance, MMs, and lifetime.

## 6    Summary

Northrop Grumman has designed an observatory flight system for the *SALTUS* NASA APEX mission concept that draws heavily from LEOStar-3 and GEOStar-3 spacecraft product lines, both of which have significant flight heritage in a variety of missions, including NASA science missions. The spacecraft architecture accommodates the *SALTUS* Payload thereby delivering unprecedented spectral sensitivity and performance to probe important far infrared science from space. The *SALTUS* M1 has a 14-meter diameter and is radiatively cooled to <45 K by a two-layer sunshield at ~1,000 m$^2$ per layer. The observatory has access to two 20° fields around the ecliptic poles and can cover the entire sky in six months. The architecture is fundamentally limited by a two consumable system driven by propellant and inflatant capacities, and the design and CONOP have been optimized to meet the 5-year mission lifetime and beyond. If selected, the mission would launch in 2032.

*Code and Data Availability Statement*

Data sharing is not applicable to this article, as no new data were created or analyzed.

**Leon Harding** is a Senior Staff Mission Architect in Science and Robotic Exploration at Northrop Grumman. He received a Ph.D. in astrophysics from the University of Galway in Ireland. Before Northrop Grumman, he was faculty at Virginia Tech's National Security Institute and was Assistant Director of the Mission Systems Division. He was a Technologist at the Jet Propulsion Laboratory, and a Scientist at Caltech. He is the *SALTUS* Spacecraft Lead and Deputy Chief Observatory Architect.

**Jonathan Arenberg** is currently the Chief Mission Architect for Science and Robotic Exploration at Northrop Grumman. His work experience includes the Chandra X-ray Observatory, co-invention of the starshade and the James Webb Space Telescope. He is the Chief Architect and Project Systems Engineer for *SALTUS*. He is an Associate Fellow of the AIAA and a Fellow of SPIE.

**Benjamin Donovan** is a Principal Systems Engineer at Northrop Grumman supporting Science and Robotic Exploration programs. He received his B.S. in physics and astronomy from The University of Iowa in 2016 and his M.S. and Ph.D in astronomy and astrophysics in 2019 and 2022, respectively, from The Pennsylvania State University. He is a member of the Spacecraft Systems Engineering team on *SALTUS*.

**Dave Oberg** is a Program Manager with Northrop Grumman, in the Science and Robotic Exploration unit. In this position he works with scientists on initial concept development and then leads the team through design, production, and launch. He holds a Bachelors and Masters degree in Aeronautics and Astronautics from the Massachusetts Institute of Technology.

**Ryan Goold** is a Staff Structural Engineer at Northrop Grumman and is responsible for the modelling of vibration disturbance sources and their propagations within a spacecraft to accurately predict the on-orbit dynamics for various satellite systems. He received a B.S. in Mechanical Engineering from the University of Missouri, and a M.S. in Mechanical Engineering, System Dynamics and Controls, from Arizona State University. His interests include on-orbit, multi-body dynamics and vibration isolation systems.

**Bob Chang** is currently in the Business Development team of the Northrop Grumman Deployables Operating unit which is responsible for large space deployables such as booms and AstroMesh reflector assemblies. Prior to this current role, Bob spent 10 years as a mechanisms engineer with contributions to various programs such as GOES, Osiris-REX, SWOT, Parker Solar Probe, Bepi Colombo and JUICE.

**Christopher Walker** is a Professor of Astronomy, Optical Sciences, Electrical Engineering, and Aerospace Engineering at UArizona. He has published numerous papers on star formation and terahertz technology, served as dissertation director to sixteen Ph.D. students, and been a Topical Editor for IEEE Transactions on Terahertz Science and Technology. Prof. Walker has been PI of numerous NSF and NASA projects and authored two textbooks; Terahertz Astronomy and Investigating Life in the Universe.

**Dana Turse** is a Space Systems Architect at Redwire. She studied Mechanical Engineering at Colorado State University and has over 20 years of experience in advanced composites, structures, and mechanisms. She has served as Chief Engineer on missions involving masts, solar arrays, tensioned membranes, and antennas, and holds over 15 patents for the associated technologies. Ms. Turse is a Co-Investigator on NASA's Solar Cruiser Mission, responsible for engineering the deployable solar sail system.

**Jim Moore** has 35 years' experience in design and analysis of advanced aerospace hardware. His areas of expertise include large lightweight space structures, membrane mechanics, thermal analysis, and optical design. He served as Chief Engineer on NeXolve's subcontract to design, manufacture, and test the James Webb Sun Shield Membrane Sub-System. He has authored and co-authored 34 technical papers and a chapter in the AIAA Gossamer Spacecraft Design Textbook.

**James C. Pearson Jr. (Jim)** leads a team of engineers, physicists, and manufacturing technicians at NeXolve who design and develop large deployable space structures, many made from thin films. These film products include sun shields, solar sails, antennae, photovoltaic systems, and spacecraft covers. Exemplar NeXolve legacy programs include the James Webb Space Observatory, NEAScout, Nanosail-D, and Solar Cruiser. He received a BS in Aerospace Engineering from Auburn University.

**John N. Kidd Jr.** is the Chief Aerospace Engineer at Ascending Node Technologies, LLC where he is responsible for conducting mission feasibility analysis and developing aerospace simulation capabilities within Spaceline, a web-based mission design and analysis platform. John studied Aerospace Engineer (B.S.) and Systems Engineering (M.S.) at the University of Arizona, where he launched his career on NASA's OSIRIS-REx Asteroid Sample Return Mission. John's interests include trajectory design and optimization, mission planning, and software engineering.

**Zach Lung** is a Space Program Strategist, Business Development Manager, and Scheduling Lead at DA2 Ventures. He received his BS in business from Colorado State University in 2016. His professional background includes mission planning, proposal management, and technical writing for astrophysics, heliophysics, and ground-based Astronomy observatories, including work on the Giant Magellan Telescope (GMT) as part of the US Extremely Large Telescope Program (US-ELTP).

**Dave Lung** has over 40 years of aerospace industry experience in a wide variety of air and space technologies and systems. He is currently Founder/President of DA2 Ventures, a consulting firm specializing in tech transfer, program planning, capture management, and proposal management. His education includes BS Aerospace Engineering, University of Minnesota (1985), MS Reliability Engineering, University of Arizona (1987), MBA University of Texas (1988), and post-graduate studies in Systems Engineering (USAF-AFIT). Officer, USAF (1984-1992).

## Caption List

**Fig. 1** The *SALTUS* observatory. The partners listed are: University of Arizona (UA; PI-institution and responsible for HiRX), Space Dynamics Laboratory (SDL; Payload/CCM manager), SRON Netherlands Institute for Space Research (SAFARI-*Lite*), L'Garde (M1 primary reflector), NeXolve and Redwire Space (Sunshield Module). Northrop Grumman provides mission management, SMA, and resources to the MOC, and is responsible for the spacecraft and observatory architecture at large, including the composite instrument deck, CIM, ICS, deployable boom, and truss.

**Fig. 2 (a)** *SALTUS* Sun-Earth halo L2 orbit. The orbit has been sized such that Earth eclipses are avoided with minimal ΔV. The Moon's orbit is shown for scale. **(b)** *SALTUS* mission operations architecture, jointly supported by Northrop Grumman and University of Arizona's Mission Operations Center (MOC), Science Operations Center (SOC), and the Lunar Exploration Ground Sites (LEGS).

**Fig. 3** *SALTUS* launch into a Sun-Earth L2 halo orbit annotaed with maneuvers and sequence of activities to prepare for science operations. The Moon's orbit is shown for scale.

**Fig. 4** *SALTUS* flight system block diagram. The legend indicates selectively redundant or internally redundant components. LEOStar-3 avionics (IEM) are coupled with a Payload Interface Electronics (PIE) module to meet *SALTUS* science and mission requirements. Both the inflation control system (ICS) and Sunshield Module (SM) are captured under the Payload, as well as the Cold Corrector Module (CCM), Warm Instrument Electronics Module (WIM), and Primary Reflector Module (PRM). The ICS includes four tanks (helium gas) mounted externally to the GEOStar-3 primary structure. The CCM resides in the cryogenic environment (M1 side, <45 K), whereas the WIM resides inside the bus structure in the warm environment for thermal considerations (spacecraft side, ~310 K).

**Fig. 5 (a)** *SALTUS* deployed configuration. The "warm side" is Earth and Sun facing. M1 boreslight is ~90° (±20° pitch, ±5° roll) to the Sun line, and the Z axis can rotate 360° around this line. **(b)** *SALTUS* launch configuration. The observatory fits in the static envelope of the LV fairing. We also show observatory Center of Gravity (CG) w.r.t. the spacecraft origin at the separation plane ("Sep plane"), and inertia in stowed and deployed configurations. **(c)** *SALTUS* stowed configuration. Sunshield layers are colored to indicate layer 1 (blue, cryo side) and layer 2 (red, warm side).

**Fig. 6** *SALTUS* cryogenic heat map. Acronyms in figure: ADR – adiabatic demagnetization refirgerator; CCM – cold corrector module; LNA – low noise amplifier; LO – local oscilator; QCL – quantum cascade lase; WIM – warm instrument module.

**Fig. 7 (a)-(d)** *SALTUS* AstroMesh architecture showing the stowed and deployed configurations, as well as two key steps in the truss deployment: release & bloom, and powered deployment. Note, the golden mesh reflective surface shown in this figure is illiustrative only; the *SALTUS* M1 inflatable membrane will instead occupy this volume.

**Fig. 8 (a)-(f)** *SALTUS* Primary Reflector Module (PRM) deployement sequence. The solar array and HGA are deployed prior to PRM deployment. The sunshield is deployed following PRM deployment. This sequence is important to avoid a cryogenic PRM deployment (which is a more complex proposition). Note, each hinge will not be required to fully deploy or latch before the next hinge actuator in the sequence initiates its motion. This strategy allows for positioning optimizations, clearance optimizations, moment of inertia impacts, etc., and will be refined in Phase A. STEP 6 (f) (i-iv) shows the four actuator boom locations.

**Fig. 9 (a)** *SALTUS* observatory isolated versus hard-mounted RWAs induced vibration RMS (1σ) translational displacements of M1. **(b)** *SALTUS* observatory isolated versus hard-mounted RWAs induced vibration RMS (1σ) rotational displacements of M1.

**Fig. 10** Rotational displacement of M1 as a function of time for a range of structural damping values.

**Fig. 11** *SALTUS* sunshield layer features and layout.

**Fig. 12** *SALTUS* sunshield stowed and deployed configurations. The stowed configuration shows a zoomed in image of the Sunshield Module (SM) which sits outside, and attaches to the CIM. The smaller ~48.5 m × ~19.5 m layer 1 deploys first, followed by the larger ~50 m × ~20 m layer 2, which avoids a cryogenic deployment of the second layer to deploy. Layer 2 will be the last major deployment of the *SALTUS* observatory.

**Fig. 13** *SALTUS* sunshield deployment sequence. (a) Stowed configuration showing each sunshield layer and deployment system location. The stowed layers, deployment system, and TRAC masts are co-spooled on a stowage hub. (b) TRAC masts are deployed using a single brushless DC motor mechanism. (c) The deployment mechanism continues to unfurl the layer. (d) Tension is driven into the layer on full deployment. The corners of each layer are mechanically

attached to the distal ends of each mast. (e) STEPS 1-4 are repeated for the second layer ensuring a non-cryogenic deployment for both layers. Once both layers are deployed, all major observatory deployments are complete and the cryogenic thermal environment begins to stabilize to <45 K.

**Fig. 14 (a)** Single deployed quadrant of the Solar Cruiser sail system utilizing 30 m TRAC masts with a side dimension of 40 m. **(b)** Membrane manufacturing facility at NeXolve which is capable of producing sunshield layers to meet the *SALTUS* required area. **(c)** A four-mast TRAC deployment system that has been matured to TRL-6.

**Fig. 15** *SALTUS* maximum allowable settle time to achieve a given observing effiency as a function of observation time. For shorter observations, the mean efficiency is greater than science mission requirements. Given the requirement of $\eta$=0.60, the mean observation time can be as short as 1h (current mean observation time is 5h CBE) and *SALTUS* will still meet this requirement.

**Fig. 16** *SALTUS* momentum unload interval in hours as a function of observatory pitch and roll at **(a)** BOL and **(b)** EOL over the *SALTUS* FOR. Each number in a colored cell coresponds to the unload interval in units of hours. The smaller the number, the shorter the unload interval, which indicates greater momentum accumulation in the RWAs at that attitude. Intervals increase with mission lifetime as propellant is used and CG migrates closer to CP. The purple boxes in (a) and (b) highlight the *SALTUS* FOR defined by ±20º of pitch, and ±5º of roll, and show a BOL minimum unload interval of 4.71h, and an EOL minimum unloaod interval of 6.59h.

**Table 1** Summary overview of *SALTUS* mission requirements.

**Table 2** *SALTUS* ΔV budget.

**Table 3** *SALTUS* mass budget summary.

**Table 4** *SALTUS* data budget summary.

**Table 5** *SALTUS* power budget summary. The driving steady-state power mode is a SAFARI-*Lite* Science observation with HiRX in stand-by with a simultaneous Ka-Band science data downlink (DL). **(1)** With a maximum MU duration of ~4 minutes and considering MU and slew battery recharge. This sizing case assumes a nominal observation and a continuous 5h science data DL. **(2)** The nominal DL cadence of 7h per week; Science + DL is used as a conservative case. Assuming return to an observation without DL, the solar array has additional margin when considering battery recharge demand and correcting MU duration. **(3)** We assume MU and slew modes draw power from the battery to avoid sizing the solar array to infrequent modes (4x per day). These modes could both be powered by the solar array.

**Table 6** *SALTUS* link budget summary. (1) C + RS represents Convolutional + RS (255, 223). (2) Margin for TT&C link maintained at minimum 6 dB for ranging-induced losses. All other links maintained at minimum 3 dB.

**Table 7** *SALTUS* Flight System margins summary. (1) LV performance capability to *SALTUS* orbit. (2) Includes full inflatant load of 200 kg. (3) Maximum dry mass assuming a full propellant tank.

**Table 8** *SALTUS* AstroMesh truss and boom flight heritage. (1) Technical details redacted due to proprietary information under NDA. Included here to highlight existing AstroMesh heritage at apertures greater than *SALTUS*. (2) The main change for *SALTUS* is a scaling of the truss for 12-14 m. (3) Demonstrated in aperture size range. The *SALTUS* boom will be scaled for *f/D*. (4) AstroMesh is Electrostatic Damage (ESD) compliant and verified Passive Intermodulation (PIM)-free by design. (5) Driven by aperture, not mission unique. (6) Includes 20% contingency.

**Table 9** *SALTUS* observatory deployed structural mode estimates. The coupling of the low frequency first mode of the deployed sunshield layer with the low mass of each layer is not anticipated to have substantial effects on pointing performance. However, this will be included in a more detailed pointing vibration analysis in Phase A.

**Table 10** *SALTUS* Sunshield Module heritage. The table contains heritage tensioned-membrane systems developed by NeXolve and Redwire Space. Note, *SALTUS* material thickness is driven by micrometeoroid considerations. The Solar Cruiser development is highlighted in green beside *SALTUS* (blue) as the most recent flight heritage. (1) Layer 2 area of ~1,000 m$^2$ from ~50 m × ~20 m dimension. Layer 1 area estimated at ~931 m$^2$ from ~48.5 m × ~19.2 m.

**Table 11** *SALTUS* Sunshield Module component masses. The total MEV mass is captured under "Payload" (Table 3).

**Table 12** *SALTUS* on-sky efficiency key & driving variables. (1) 7h per week distributed per CONOP. (2) Expected to occur at shortest interval, worst case BOL attitude. (3) From LEOStar-3 flight actuals. (4) CBE observation time is 5 hours; 4 hours used for added buffer. Therefore, assume 6 observations per day, each requiring a slew to target. (5) Observatory overhead to capture uncertainties in the design process, which we add as an additional $\gamma$=25% to 12.3%.